\documentclass[aps,twocolumn,superscriptaddress,groupedaddress,nofootinbib]{revtex4-1}  % 

\usepackage{ifpdf}

\usepackage{graphicx}  
\usepackage{dcolumn}   
\usepackage{bm}        
\usepackage{amssymb}

\usepackage{amsmath,amssymb}

\usepackage{verbatim}
\usepackage{float}
\usepackage{subcaption} 
\usepackage{amsfonts}
\usepackage[inline]{enumitem}
\newlist{inlinelist}{enumerate*}{1}
\setlist*[inlinelist,1]{
  label=(\arabic*),
}
\usepackage{mathtools}

\usepackage[usenames,dvipsnames]{xcolor}
\definecolor{darkerblue}{rgb}{0.2,0.2,0.5}

\usepackage{tikz}

\usepackage{afterpage}

\newcommand{\slashed}[1]{\displaystyle{\not}#1}

\newcommand{\bear}{\begin{array}}
\newcommand{\ear}{\end{array}}

\newcommand{\beq}{\begin{eqnarray}}
\newcommand{\eeq}{\end{eqnarray}}
\newcommand{\beqa}{\begin{eqnarray}}
\newcommand{\eeqa}{\end{eqnarray}}

\def\OMIT#1{{}}
\newcommand{\lsim}{\mathrel{\rlap{\lower4pt\hbox{\hskip1pt$\sim$}}
    \raise1pt\hbox{$<$}}}        
\newcommand{\gsim}{\mathrel{\rlap{\lower4pt\hbox{\hskip1pt$\sim$}}
    \raise1pt\hbox{$>$}}}

\newcommand{\LL}{\mathcal{L}}

\newcommand{\OO}{\mathcal{O}}

\newcommand{\MET}{\slashed{E}_T}

\begin{document}

\title{Jet Observables and Stops at 100 TeV Collider}
\author{JiJi Fan} 
\author{Prerit Jaiswal}
\author{John Shing Chau Leung}
\affiliation{Department of Physics, Brown University, Providence, RI, 02912, USA}
\date{\today}

\begin{abstract}
A future proton-proton collider with center of mass energy around 100 TeV will have a remarkable capacity to discover massive new particles and continue exploring weak scale naturalness. In this work we will study its sensitivity to two stop simplified models as further examples of its potential power: pair production of stops that decay to tops or bottoms and higgsinos; and stops that are either pair produced or produced together with a gluino and then cascade down through gluinos to the lightest superpartner (LSP). In both simplified models, super-boosted tops or bottoms with transverse momentum of order TeV will be produced abundantly and call for new strategies to identify them. We will apply a set of simple jet observables, including track-based jet mass, $N$-subjettiness and mass-drop, to tag the boosted hadronic or leptonic decaying objects and suppress the Standard Model as well as possible SUSY backgrounds. Assuming 10\% systematic uncertainties, the future 100 TeV collider can discover (exclude) stops with masses up to 6 (7) TeV with 3 ab$^{-1}$ of integrated luminosity if the stops decay to higgsinos. If the stops decay through gluinos to LSPs, due to additional SUSY backgrounds from gluino pair production, a higher luminosity of about 30 ab$^{-1}$ is needed to discover stops up to 6 TeV. 
We will also discuss how to use jet observables to distinguish simplified models with different types of LSPs. The boosted top or bottom tagging strategies developed in this paper could also be used in other searches at a 100 TeV collider. For example, the strategy could help discover gluino pair production with gluino mass close to 11 TeV with 3 ab$^{-1}$ of integrated luminosity. 
\end{abstract}

\maketitle

%%%%%%%%%%%%%%%%%%%%%%%%%%%%%%%%%%%%%%%%%%%%%%
\section{Introduction}
\label{sec:introduction}
%%%%%%%%%%%%%%%%%%%%%%%%%%%%%%%%%%%%%%%%%%%%%%

Collider experiments have been the most powerful probe to reveal the nature of the smallest possible distance scale in particle physics. While currently the Large Hadron Collider (LHC) is still busy exploring the TeV scale, there has been a growing effort in planning for future hadron colliders to take the baton from the LHC in hunting for new physics beyond the Standard Model (SM)~\cite{Avetisyan:2013onh, Richter:2014pga, Rizzo:2015yha, Benedikt:2015poa, Tang:2015qga, Arkani-Hamed:2015vfh, CEPC-SPPCStudyGroup:2015csa, Mangano:2016jyj, Golling:2016gvc, Contino:2016spe}. So far the most discussed future hadron collider scenario is a circular 100 TeV proton-proton machine. It has been demonstrated that such a machine can push the testable energy frontier by roughly one order of magnitude and could discover colored particles with masses near 10 TeV \cite{Borschensky:2014cia,Stolarski:2013msa, Cohen:2013xda, Jung:2013zya, Cohen:2014hxa,Fowlie:2014awa,Ellis:2015xba,Antusch:2016nak,diCortona:2016fsn} as well as electroweak particles with masses near 1 TeV \cite{Low:2014cba, Cirelli:2014dsa, Hook:2014rka, Gori:2014oua, Bramante:2014tba, Alves:2014cda,  Hajer:2015gka, Bramante:2015una, Ismail:2016zby, Mahbubani:2017gjh, Fukuda:2017jmk, Kobakhidze:2016mdx}.

One of the most-motivated new physics targets at hadron colliders is the top partners, for example, stops in the supersymmetric (SUSY) scenarios. The mass scale of stops is an indication of the fine-tuning level of electroweak symmetry breaking in SUSY~\cite{Barbieri:1987fn, Dimopoulos:1995mi, Pomarol:1995xc, Cohen:1996vb, Kitano:2006gv, Perelstein:2007nx, Papucci:2011wy, Brust:2011tb}. So far only the simplest possible stop decay, $\tilde{t}  \to t + \tilde{\chi}_1^0$ with $\tilde{\chi}_1^0$ being the lightest neutralino has been studied at a 100 TeV collider~\cite{Cohen:2014hxa}.

In this article, we will investigate reach of a 100 TeV collider for stops in two new stop simplified models with more complicated stop decay chains and final state topologies. In the stop-higgsino model\footnote{This model is also considered in Ref.~\cite{Graesser:2012qy} in the context of the LHC.} ($\tilde{t}-\tilde{H}$ model), the higgsino multiplet is at the bottom of the SUSY spectrum. Right-handed stops will decay to both neutral and charged higgsinos, which are nearly degenerate in mass, with about equal probabilities. In the $\tilde{t}-\tilde{g}-\tilde{\chi}_1^0$ model, the gluino is lighter than the stops. The stops will cascade down to the the lightest neutralino (which we take to be bino) through the gluino and produce multiple tops. In this case, stop-gluino associated production could be as important as stop pair production.

In the stop searches, one generic challenge is that SM particles, in particular, tops produced from decays of the heavy stops would be hyper-boosted with transverse momentum of order TeV and above. Their subsequent decay products would be collimated into a small cone with angular size comparable to or even smaller than a calorimeter cell. This makes the standard tagging procedure used at the LHC not directly applicable. In Ref.~\cite{Cohen:2014hxa}, it is suggested that leptonic-decaying tops could be identified by tagging a hard muon inside the jet at a 100 TeV collider. To study the more complicated stop decay topologies, we need to go beyond the simple muon tagging strategy and tag hadronic-decaying tops to improve the reach. We will develop boosted top and $b$ jet tagging strategies based on several jet observables such as the track-based observables discussed in Ref.~\cite{Larkoski:2015yqa} to suppress both the SM and SUSY backgrounds.

The paper is organized as follows: in Sec.~\ref{sec:models}, we present details of the two stop simplified models. In Sec.~\ref{sec:observables}, we discuss the jet finding algorithms and demonstrate the discriminating powers of several jet observables we used in the analyses. In Sec.~\ref{sec:analysis1}, we present the analysis for the $\tilde{t}-\tilde{H}$ model and its results. In Sec.~\ref{sec:analysis2}, we present the analysis for the $\tilde{t}-\tilde{g}-\tilde{\chi}_1^0$ model and its results. We will conclude and discuss possible future directions in Sec.~\ref{sec:con}.

%%%%%%%%%%%%%%%%%%%%%%%%%%%%%%%%%%%%%%%%%%%%%%
\section{Simplified Models}
\label{sec:models}
%%%%%%%%%%%%%%%%%%%%%%%%%%%%%%%%%%%%%%%%%%%%%%
We consider two new simplified models: $\tilde{t}-\tilde{H}$ and $\tilde{t}-\tilde{g}-\tilde{\chi}^0_1$, which will be described in detail below. For simplicity, we only consider right-handed stops in the simplified models. 

\begin{figure}
\begin{subfigure}[h]{0.4\textwidth}
\vspace{-4.5in}
\begin{center}
\includegraphics{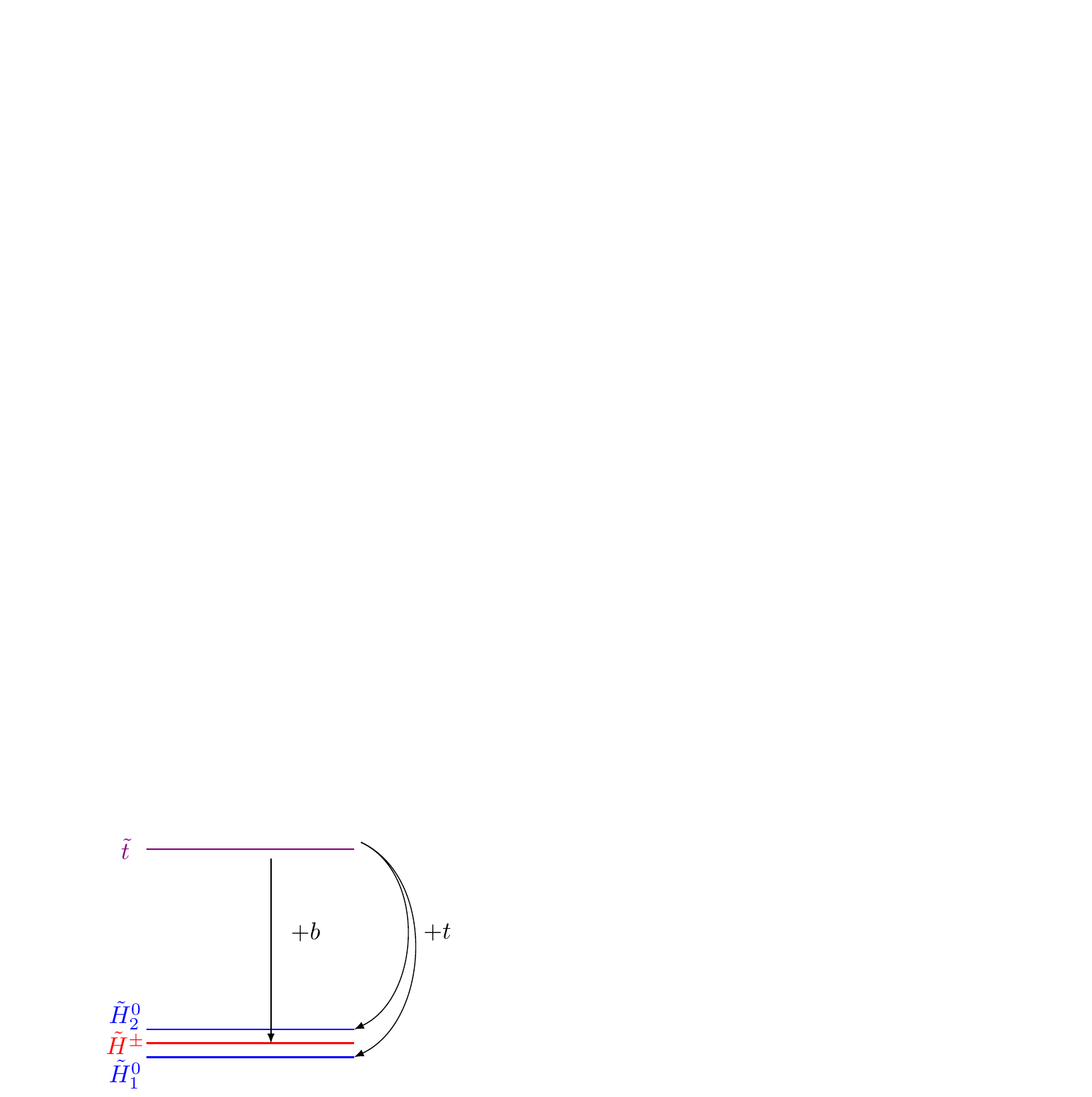}
\end{center}\vspace{-0.6cm}
\caption{Higgsino LSP}\label{fig:higgsino_spectrum}
\end{subfigure}
\begin{subfigure}[h]{0.4\textwidth}
\centering\vspace{-4.5in}
\includegraphics{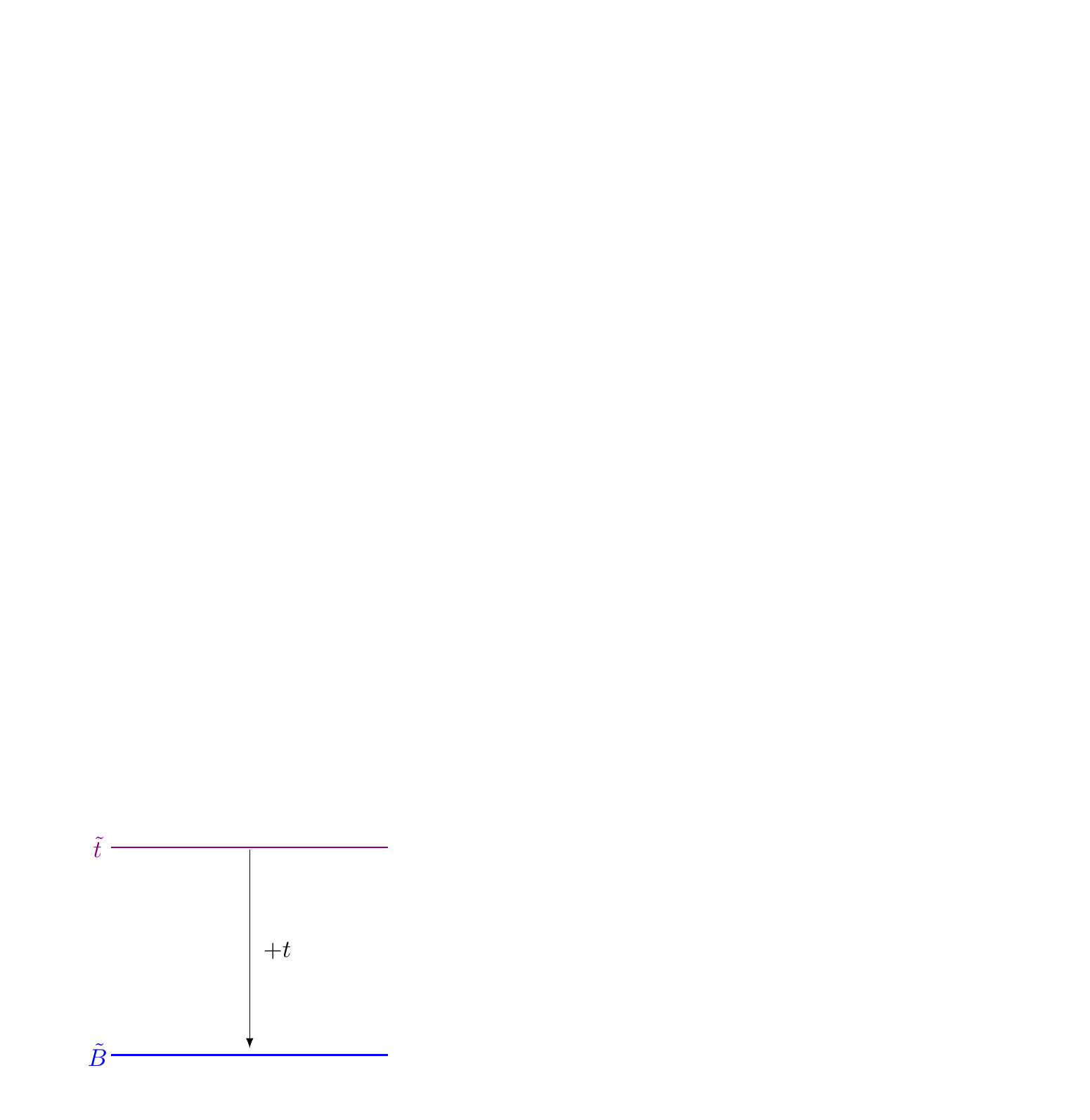}\vspace{-0.7cm}
\caption{Bino LSP}\label{fig:gaugino_spectrum}
\end{subfigure}
\begin{subfigure}[h]{0.4\textwidth}
\centering\vspace{-4.5in}
\includegraphics{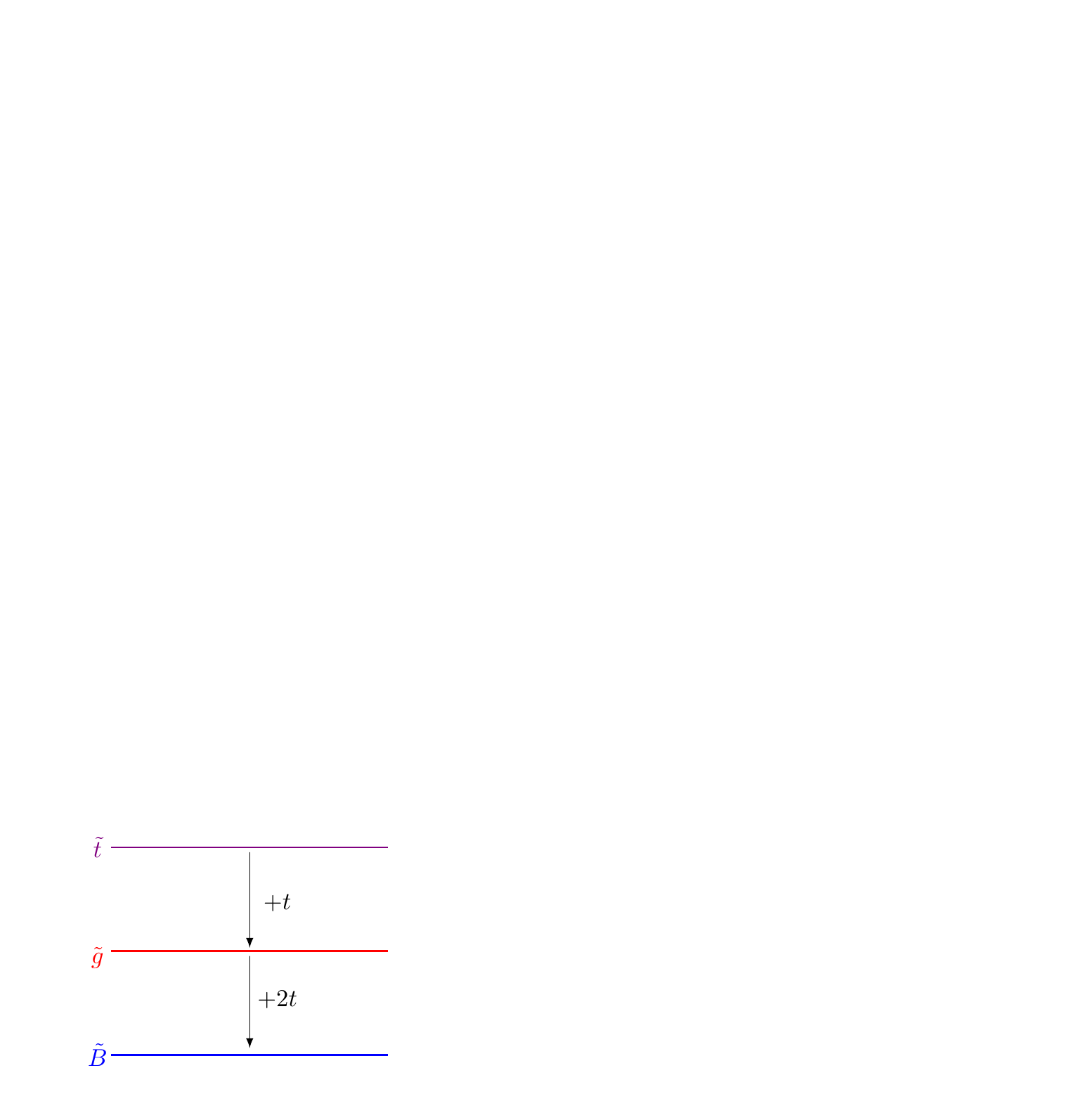}\vspace{-0.7cm}
\caption{Bino LSP with gluino NLSP}\label{fig:gluino_spectrum}
\end{subfigure}
\caption{Stop simplified models.}
\end{figure}

%%%%%%%%%%%%%%%%%%%%%%%%%%%%%%%%%%%%%%%%%%%%%%
\subsection{$\tilde{t}-\tilde{H}$ Simplified Model}
%%%%%%%%%%%%%%%%%%%%%%%%%%%%%%%%%%%%%%%%%%%%%%

 In the $\tilde{t}-\tilde{H}$ simplified model, the higgsino multiplet is at the bottom of the SUSY mass spectrum and $O$(TeV) lighter than the stops while the remaining SUSY particles are assumed to be decoupled. The neutral and charged higgsino masses are nearly degenerate, separated by only $O(\text{GeV}$), with the neutral higgsino $\tilde{H}^0_{1}$ being the lightest supersymmetric particle  (LSP) (fig.~\ref{fig:higgsino_spectrum}). In addition to studying the reach of $\tilde{t}-\tilde{H}$ model at a 100 TeV collider, we will also discuss how to distinguish it from the simplest stop simplified model with bino being the LSP (fig.~\ref{fig:gaugino_spectrum}).

\begin{figure}
    \centering
    \begin{subfigure}[!h]{0.35\textwidth}
        \includegraphics[width=\textwidth]{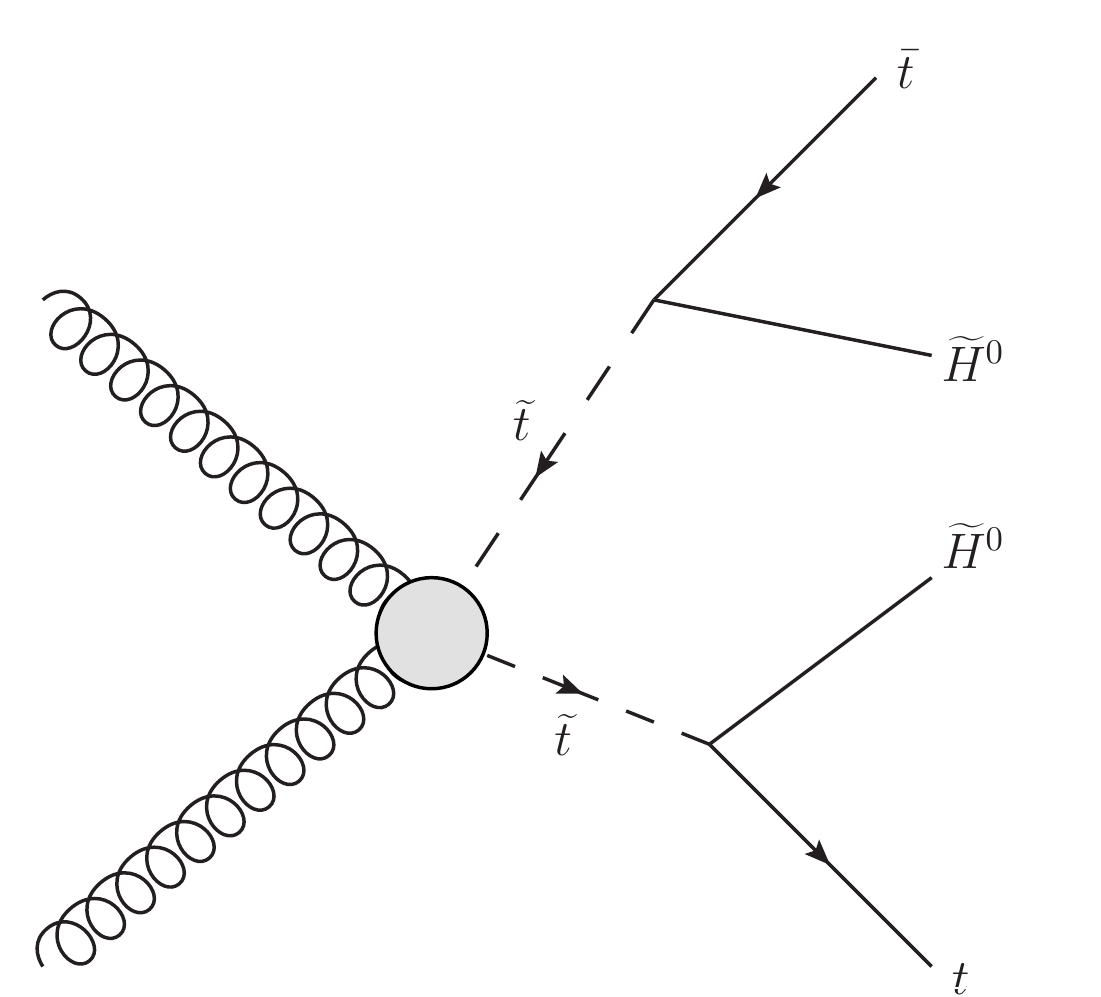}
        \caption{Stops decaying to $\tilde{H}^0$}
        \label{fig:stopneutralino}
    \end{subfigure}
    ~ \qquad
    \begin{subfigure}[!h]{0.45\textwidth}
        \includegraphics[width=\textwidth]{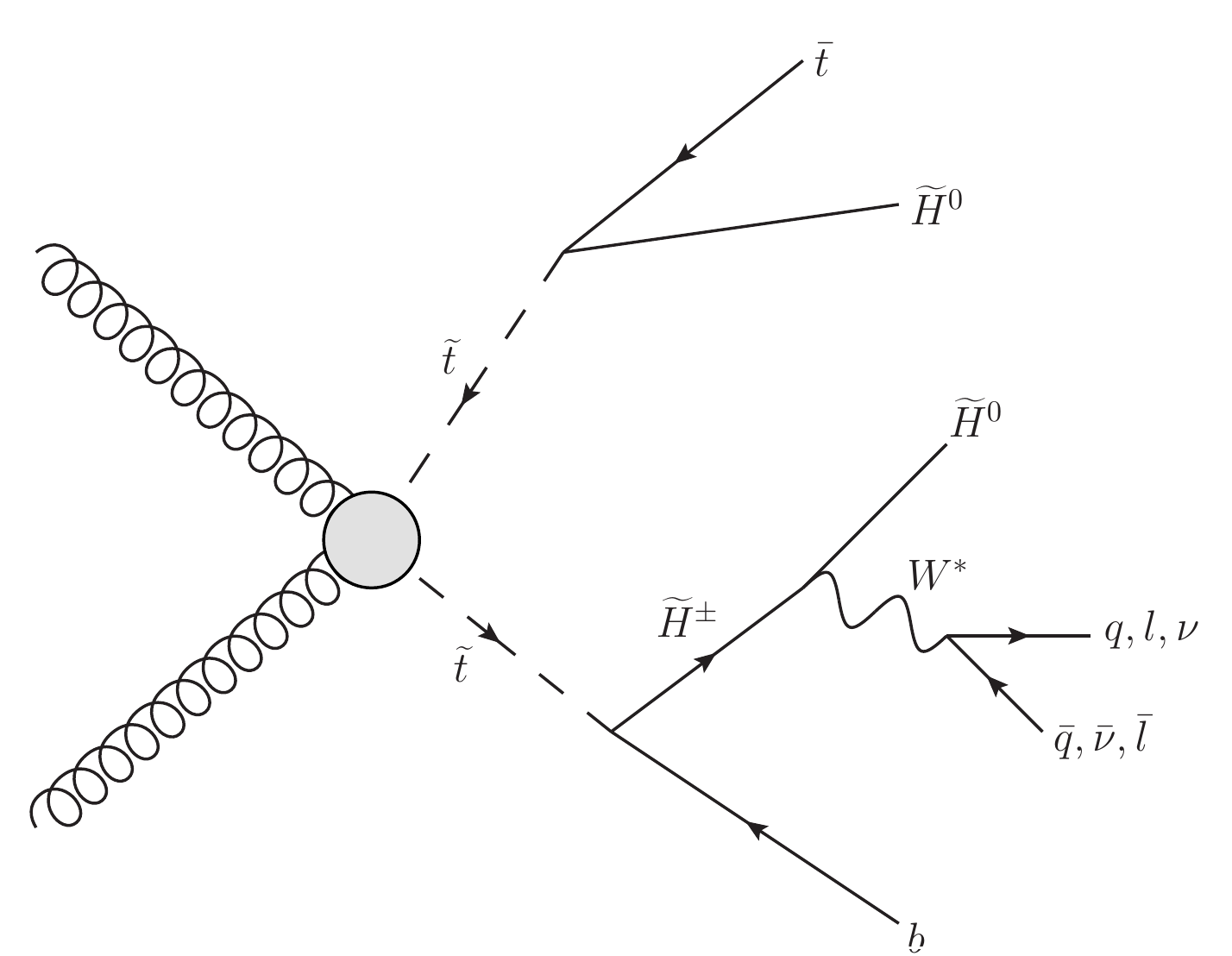}
        \caption{Stops decaying to either $\tilde{H}^0$ or $\tilde{H}^\pm$}
        \label{fig:stopmixed}
    \end{subfigure}
        ~ \qquad
    \caption{Sample Feynman diagrams for signal in the $\tilde{t}-\tilde{H}$ simplified model.}\label{fig:SUSYsignal1}
\end{figure}

There are two decay channels for stops in the $\tilde{t}-\tilde{H}$ model, each of which is equally likely:
\begin{itemize}
\item The first channel is the stop decaying to neutral higgsinos, i.e. $\tilde{t}\to t\tilde{H}^0_{2}\to tZ^*+\tilde{H}^0_1$, or $\tilde{t}\to t+\tilde{H}^0_1$ (fig.~\ref{fig:stopneutralino}). The decay emits a boosted top and the LSP, which may also be accompanied by soft particles if the stop decays to $\tilde{H}^0_2$ first. The particles from off-shell $Z^*$ decays are soft, $E \sim O( \text{GeV})$, making their measurement difficult at a hadron collider. We do not consider tagging them in this paper. 
\item The other stop decay channel is $\tilde{t}\to b\tilde{H}^{\pm}\to bW^*+\tilde{H}^0_1$ (fig.~\ref{fig:stopmixed}). The $\tilde{H}^{\pm}$ from stop decay promptly decays to the LSP and an off-shell $W^*$. Similar to the previous decay channel, SM particles resulting from $W^*$ decay are too soft to be tagged. 
\end{itemize}
The signal events will then be a mixture of $b$'s and $t$'s accompanied by missing energy. In comparison, in the $\tilde{t}-\tilde{B}$ simplified model, $\tilde{t} \to t \tilde{B}$ and the signal events contain only boosted $t$'s. 

%%%%%%%%%%%%%%%%%%%%%%%%%%%%%%%%%%%%%%%%%%%%%%
\subsection{$\tilde{t}-\tilde{g}-\tilde{\chi}_1^0$ Simplified Model}
\begin{figure}
    \centering
    \begin{subfigure}[!h]{0.35\textwidth}
        \includegraphics[width=\textwidth]{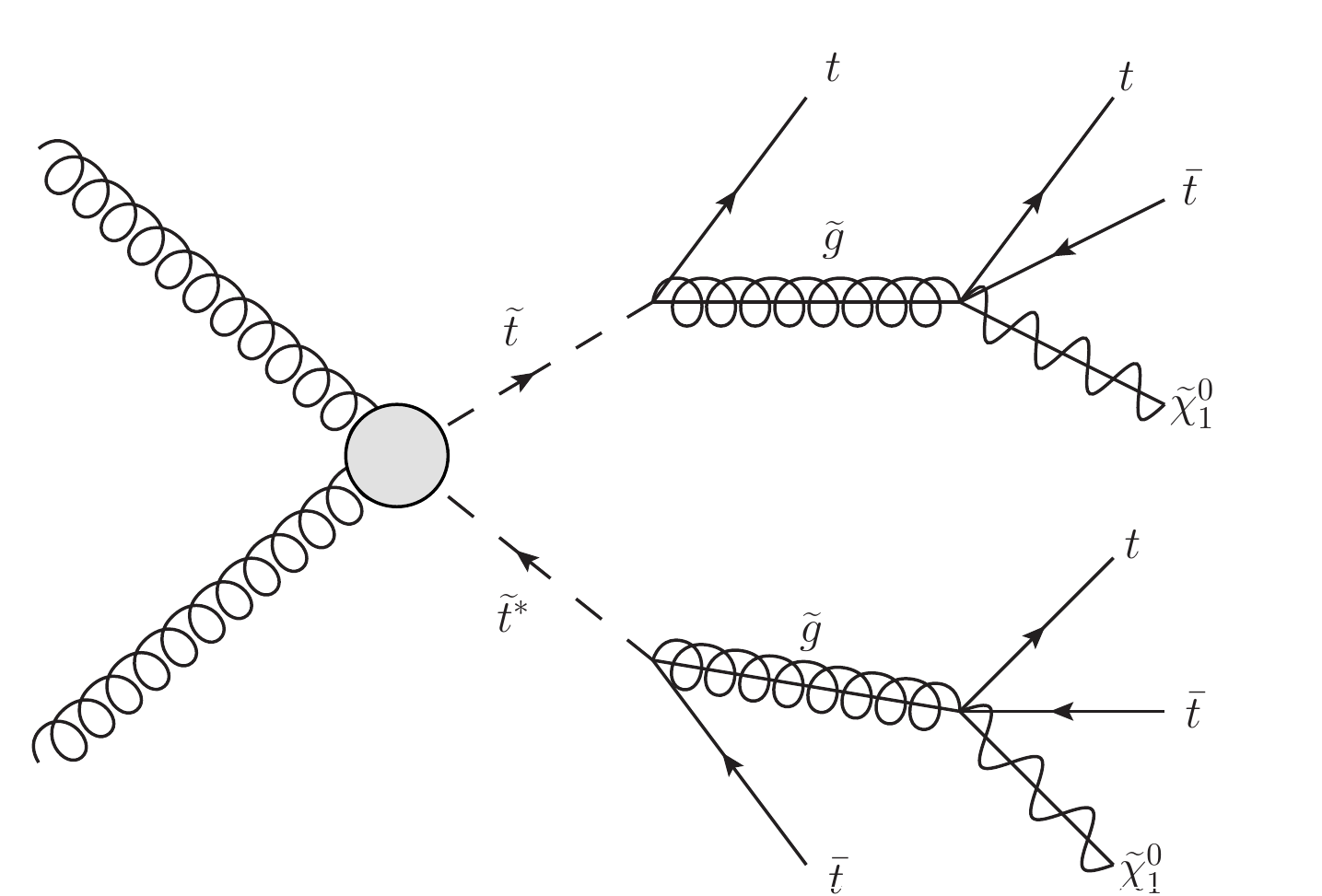}
        \caption{Stop-pair production}
        \label{fig:StSt}
    \end{subfigure}
    ~ \qquad
    \begin{subfigure}[!h]{0.35\textwidth}
        \includegraphics[width=\textwidth]{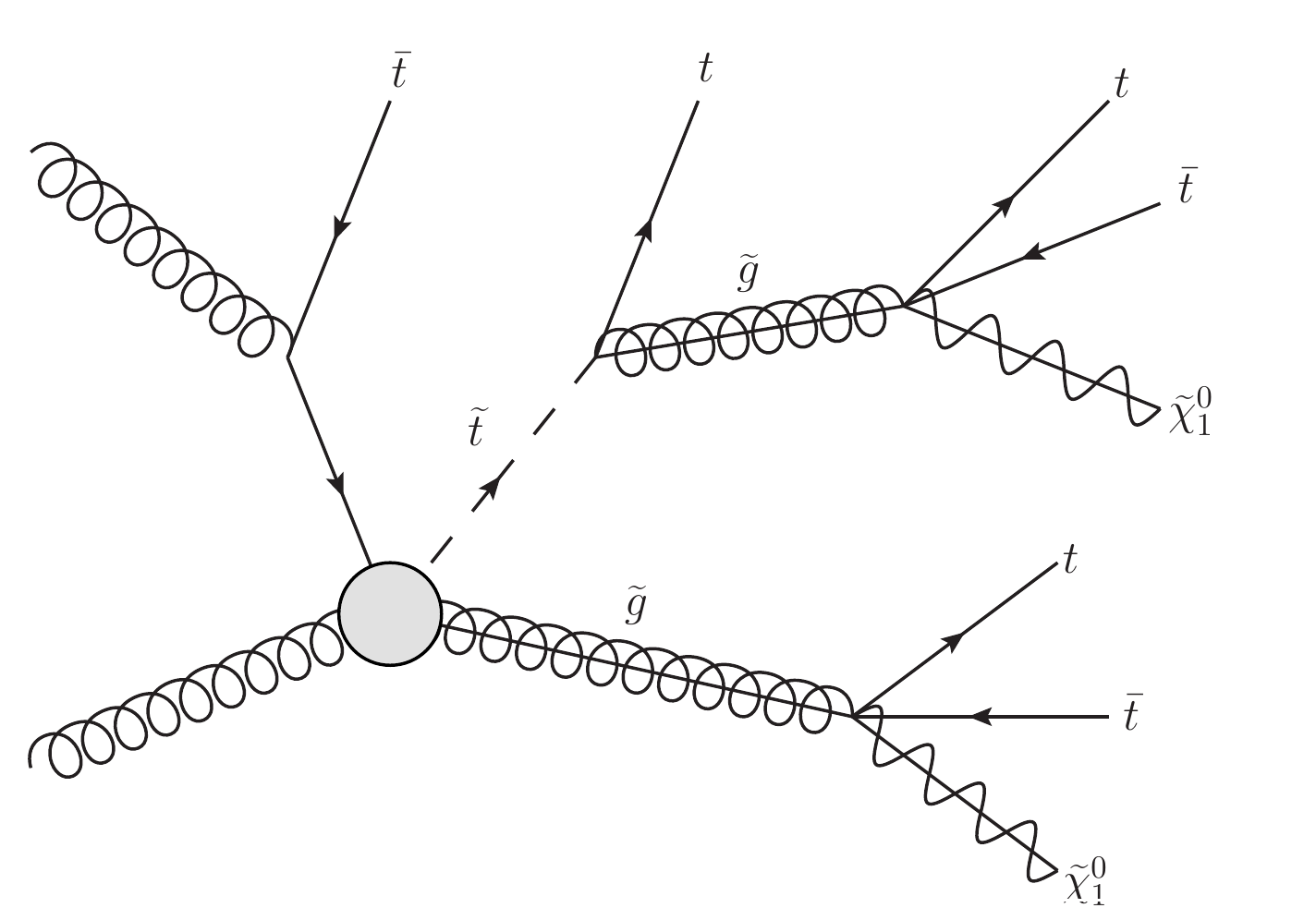}
        \caption{Stop-gluino associated production}
        \label{fig:StGl}
    \end{subfigure}
    \caption{Sample Feynman diagrams for signal in the $\tilde{t}-\tilde{g}-\tilde{\chi}_1^0$ simplified model.}\label{fig:SUSYsignal}
\end{figure}

%%%%%%%%%%%%%%%%%%%%%%%%%%%%%%%%%%%%%%%%%%%%%%
In this simplified model, we assume the three lightest SUSY particles to be stops, gluino and bino (LSP) while the remaining SUSY particles are decoupled (fig.~\ref{fig:gluino_spectrum}). Similar simplified models have been considered before in the literature for future collider searches with one major difference: previous studies assume a mass hierarchy between stops and gluino so that one of them can be decoupled from the other \cite{Cohen:2014hxa}.  In this study, however, we assume that stops and gluinos  are both $\OO(1-10 \, \mathrm{TeV})$ so that they can not be decoupled. We further assume that gluinos are lighter than stops so that the relevant decay channels are $\tilde{t} \rightarrow \tilde{g} t$ and $\tilde{g} \rightarrow t \bar{t} \tilde{\chi}_1^0$  (fig.~\ref{fig:SUSYsignal}).  Henceforth we refer to this simplified model as the $\tilde{t}-\tilde{g}-\tilde{\chi}_1^0$ model. Although this simplified model was considered in \cite{Cohen:2013xda}, the focus of that study was to estimate gluino reach at future colliders. Our goal instead is to estimate the reach of heavy stops in the context of $\tilde{t}-\tilde{g}-\tilde{\chi}_1^0$ model at future $100$ TeV collider. There are two stop production channels in this model, each characterized by 6 top quarks and missing energy in final state :
\begin{itemize}
\item Stop-pair production  : $p p \rightarrow \tilde{t} \tilde{t}^* \rightarrow t \bar{t} t \bar{t} t  \bar{t} + \slashed{E_T}$  (fig.~\ref{fig:StSt}).
\item Stop-gluino associated production : $p p \rightarrow t \tilde{t}^* \tilde{g} \rightarrow t \bar{t} t \bar{t} t  \bar{t} + \slashed{E_T}$ (fig.~\ref{fig:StGl}).
\end{itemize}

\begin{figure}
    \centering
    \begin{subfigure}[!h]{0.35\textwidth}
        \includegraphics[width=\textwidth]{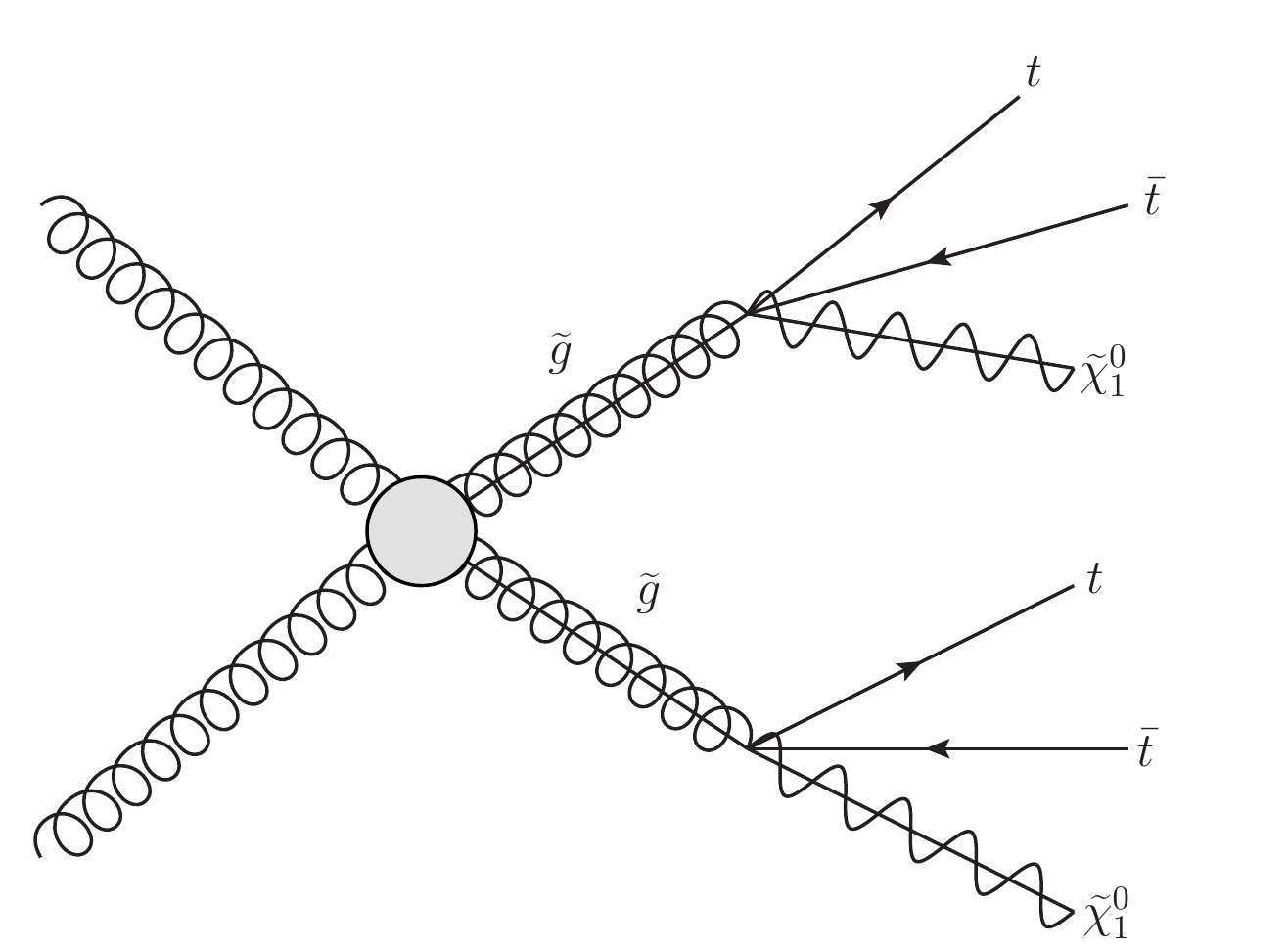}
        \caption{Gluino-pair production}
        \label{fig:GlGl}
    \end{subfigure}
    ~ \qquad
    \begin{subfigure}[!h]{0.35\textwidth}
        \includegraphics[width=\textwidth]{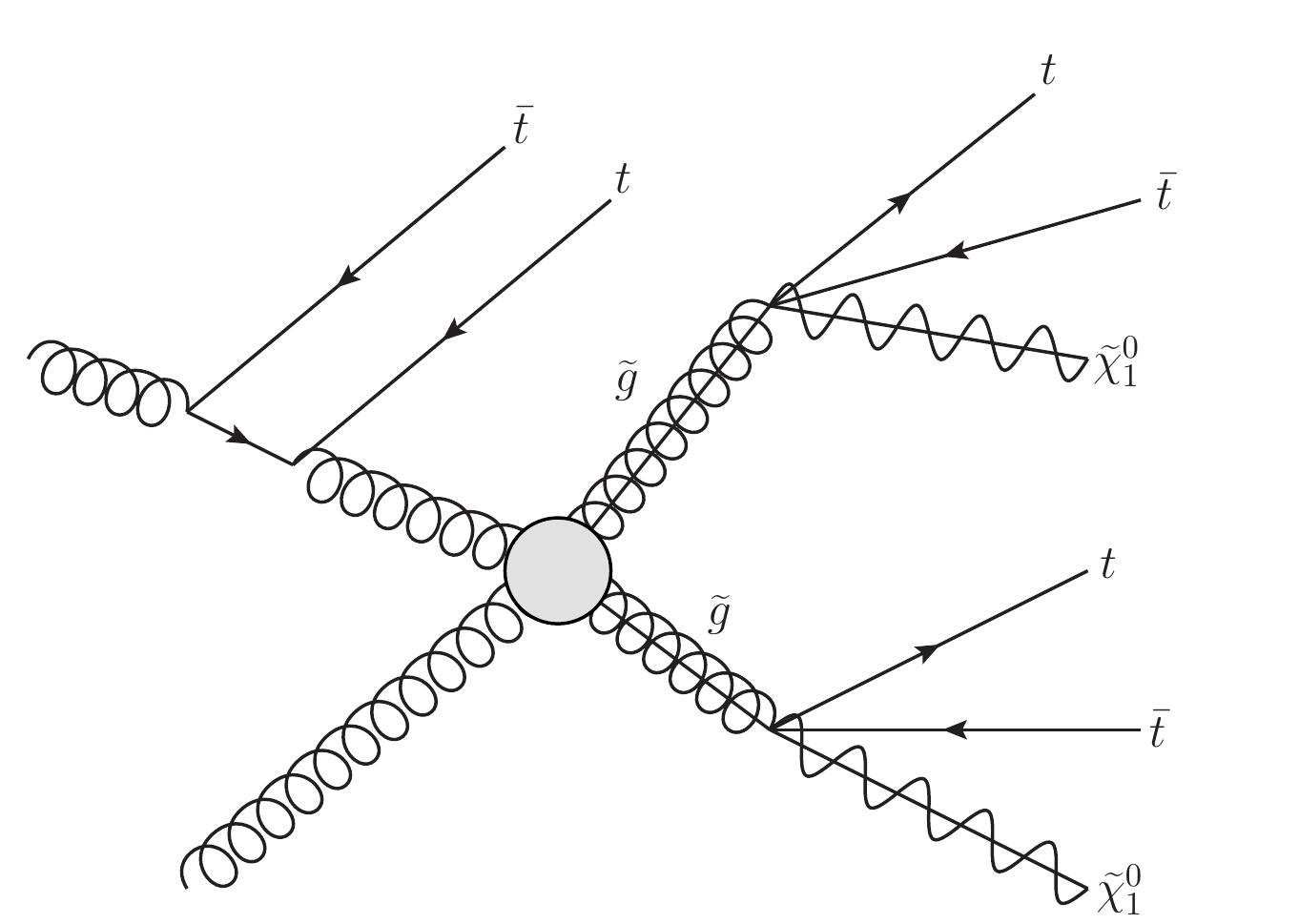}
        \caption{Gluino-pair production with tops}
        \label{fig:GlGlTops}
    \end{subfigure}
    \caption{Sample Feynman diagrams for SUSY background in the $\tilde{t}-\tilde{g}-\tilde{\chi}_1^0$ simplified model.}\label{fig:SUSYbg}
\end{figure}

Besides SM backgrounds, there is an additional important SUSY background to be considered. The mass hierarchy of $\tilde{t}-\tilde{g}-\tilde{\chi}_1^0$ model implies that the gluino must have already been discovered before the stops. Therefore, we must also consider the following gluino-pair production channels for background :

\begin{itemize}
\item Gluino-pair production : $p p \rightarrow \tilde{g} \tilde{g} \rightarrow t \bar{t} t  \bar{t} + \slashed{E_T}$ (fig.~\ref{fig:GlGl}).
\item Gluino-pair production with tops :  $p p \rightarrow \tilde{g} \tilde{g} t \bar{t} \rightarrow t \bar{t} t  \bar{t} t   \bar{t} t + \slashed{E_T}$ (fig.~\ref{fig:GlGlTops}).
\end{itemize}

%%%%%%%%%%%%%%%%%%%%%%%%%%%%%%%%%%%%%%%%%%%%%%
\section{Event Generation and Jet Observables}
\label{sec:observables}
%%%%%%%%%%%%%%%%%%%%%%%%%%%%%%%%%%%%%%%%%%%%%%%%

%%%%%%%%%%%%%%%%%%%%%%%%%%%%%%%%%%%%%%%%%%%%%%%%%%%%
\subsection{Event Generation}
%%%%%%%%%%%%%%%%%%%%%%%%%%%%%%%%%%%%%%%%%%%%%%%%%%%%

\begin{figure}
    \centering
    \begin{subfigure}[!h]{0.35\textwidth}
        \includegraphics[width=\textwidth]{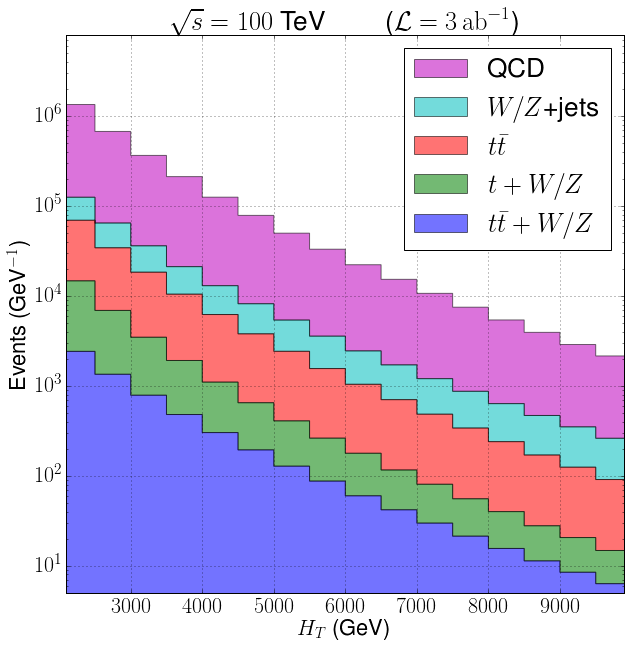}
    \end{subfigure}
    ~ \qquad
    \begin{subfigure}[!h]{0.35\textwidth}
        \includegraphics[width=\textwidth]{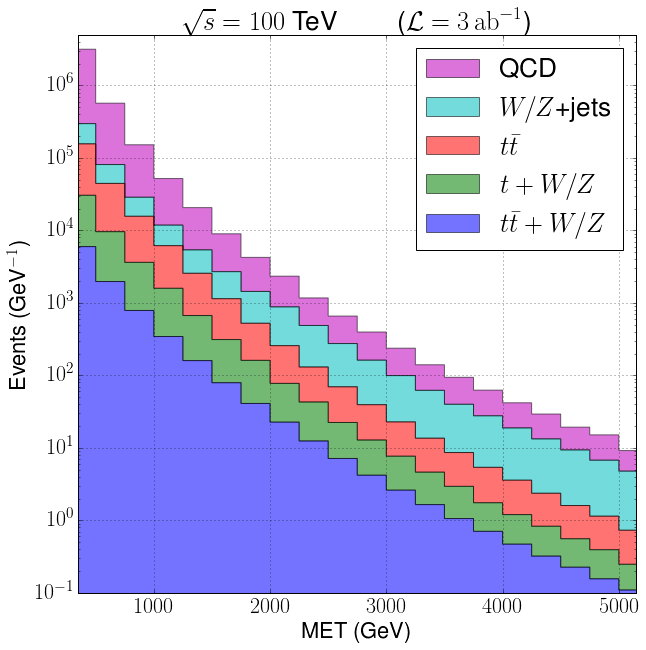}
    \end{subfigure}
    \caption{$H_T$ (left) and $\MET$ (right) distributions for SM processes.}\label{fig:SM_HT_MET}
\end{figure}

\begin{figure}
    \centering
    \begin{subfigure}[!h]{0.35\textwidth}
        \includegraphics[width=\textwidth]{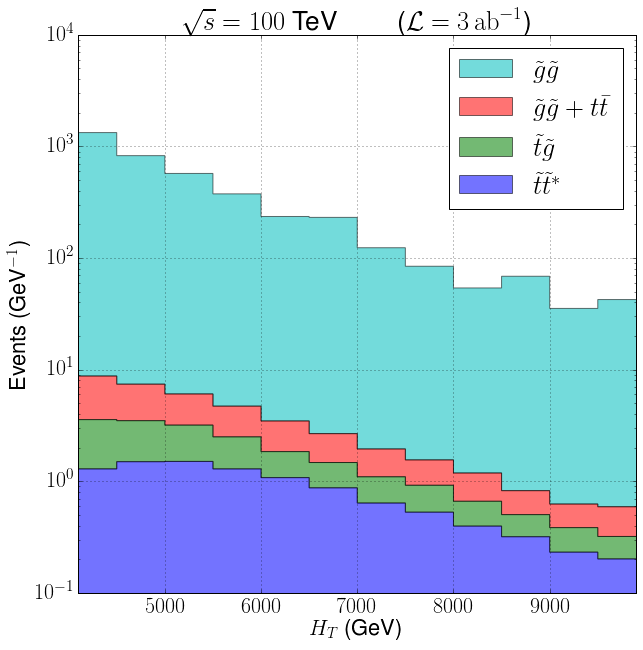}
    \end{subfigure}
    ~ \qquad
    \begin{subfigure}[!h]{0.35\textwidth}
        \includegraphics[width=\textwidth]{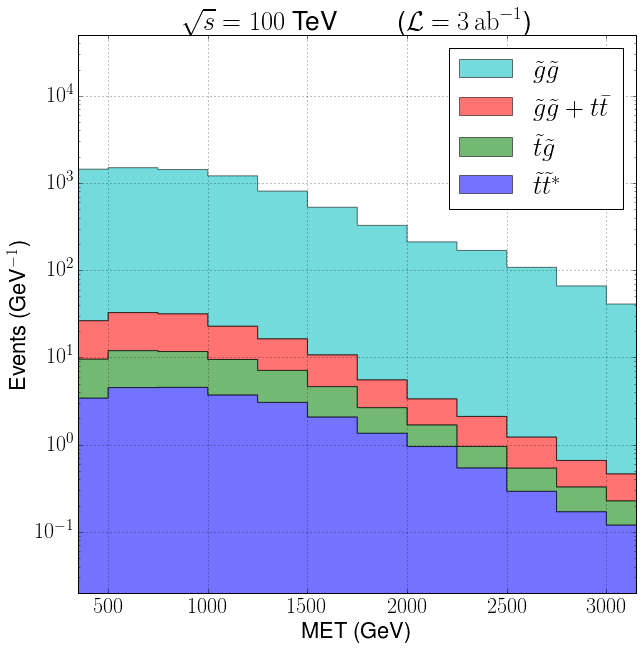}
    \end{subfigure}
    \caption{$H_T$ (left) and $\MET$ (right) distributions for SUSY processes for $m_{\tilde{t}} = 4$ TeV, $m_{\tilde{g}} = 2$ TeV and $m_{\tilde{\chi}_1^0} = 200$ GeV.}\label{fig:SUSY_HT_MET}
\end{figure}

Parton-level events were generated using {\tt MadGraph5} \cite{Alwall:2014hca}, split into four bins : $H_T \in (1.5 , 3]$ TeV, $(3, 5.5]$ TeV, $(5.5,8.5]$ TeV and $(8.5, 100]$ TeV,
followed by parton-showering and hadronization in {\tt Pythia8} \cite{Sjostrand:2007gs} and detector simulation in {\tt Delphes} \cite{deFavereau:2013fsa}. For SM background samples, additional jets were included at the parton-level\footnote{Two additional jets were included for all SM processes except $t \bar{t} +W/Z$ for which only one additional jet was included.} and then matched to parton shower using the MLM matching scheme \cite{Mangano:2006rw}.  $H_T$ and $\MET$ distributions for SM and SUSY processes are shown in fig.~\ref{fig:SM_HT_MET} and fig.~\ref{fig:SUSY_HT_MET} respectively. These distributions serve as a consistency check for correct normalization of $H_T$ bins as well as matching for SM processes.  As pre-selection cuts, events are required to have $H_T>2$ TeV and $\MET > 200$ GeV.

%%%%%%%%%%%%%%%%%%%%%%%%%%%%%%%%%%%%%
\subsection{Jet Clustering}\label{sec:jet_finding}
%%%%%%%%%%%%%%%%%%%%%%%%%%%%%%%%%%%%%
Final state hadrons and non-isolated leptons are clustered into jets using {\tt FastJet}\cite{Cacciari:2011ma} with jet radius parameter $R=0.5$ and using the anti-$k_T$ algorithm \cite{Cacciari:2008gp}. Given that both simplified models are characterized by boosted top quarks in the final state, jet substructure is a valuable tool for identifying tops. To this end, we additionally cluster fat jets with $p_T>200$ GeV using the Cambridge/Aachen (C/A) algorithm~\cite{Dokshitzer:1997in, Wobisch:1998wt} and jet radius parameter $R=1$, with the idea being that fat jets adequately capture top decay products. A well-known issue with fat jets is that the presence of  final state radiation (FSR) from top quarks and initial-state radiation (ISR)/underlying event can adversely affect the jet mass and other jet substructure properties. To mitigate this problem, Ref.~\cite{Larkoski:2015yqa} proposed scaling down the fat jet radius to $R = C m_{\text{top}}/p_T$ where $C$ is $\OO(1)$ number. The basic idea behind using dynamic radius is that the top decay products are confined to angular size of $m_{\text{top}}/p_T$ while ISR/FSR outside this cone-size is excluded.    

In our analyses, we recluster the C/A jets using the anti-$k_T$ algorithm and winner-take-all (WTA) recombination scheme \cite{Larkoski:2014bia}. In the analysis of the $\tilde{t}-\tilde{H}$ simplified model, we recluster the C/A jets with $R=600\text{ GeV}/{p_T} \approx 3.5 m_{\text{top}}/p_T$. In the $\tilde{t}-\tilde{g}-\tilde{\chi}_1^0$ simplified model, there are 6 top quarks in the final state. Occasionally, multiple top quarks are clustered into a single fat jet. To resolve this issue, we perform 
a two-step scaling down procedure. In the first step, we recluster the C/A jets with $R= (1 \,\mathrm{TeV})/p_T$ to separate multiple top quarks if any. All subjets with $p_T>500$ GeV are retained as top candidates. In the second step, each of the resulting jets are further reclustered with a smaller radius of $R= (600 \,\mathrm{GeV})/p_T$ to remove ISR/FSR. We compute jet observables which we will discuss below based on the reclustered final jets.

%%%%%%%%%%%%%%%%%%%%%%%%%%%%%%%%%%%%%%%%%%%%%%%%%%%%
\subsection{Jet Mass}
\label{sec:jetmass}
%%%%%%%%%%%%%%%%%%%%%%%%%%%%%%%%%%%%%%%%%%%%%%%%%%%%

\begin{figure}
    \centering
    \begin{subfigure}[!h]{0.42\textwidth}
        \includegraphics[width=\textwidth]{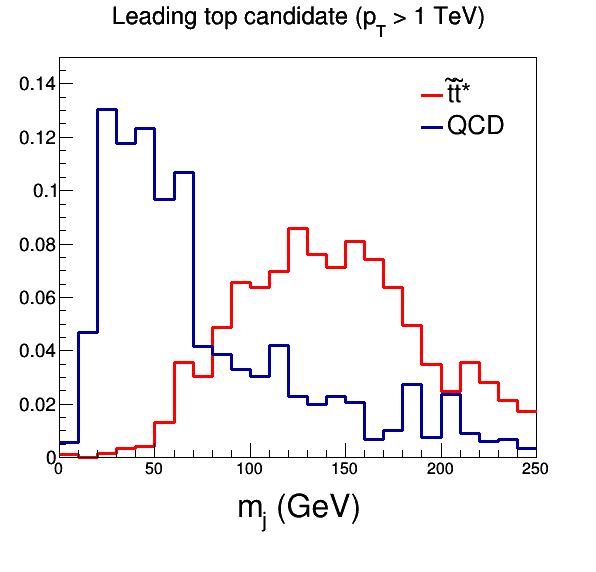}
        	\vspace{-0.8cm}
        \caption{Leptonically decaying  top candidate}
        \label{fig:mass_leptonic}
        \vspace{0.7cm}
    \end{subfigure}
    ~ \qquad
    \begin{subfigure}[!h]{0.42\textwidth}
        \includegraphics[width=\textwidth]{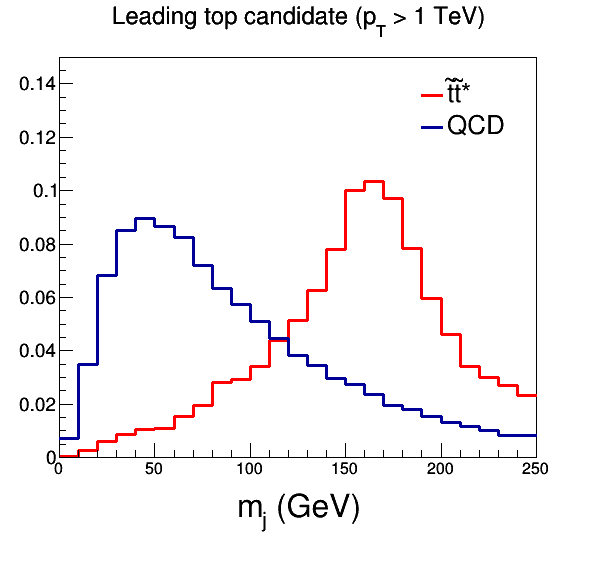}
        \vspace{-0.8cm}
        \caption{Hadronically decaying top candidate}
        \label{fig:mass_hadronic}
    \end{subfigure}
    \caption{Jet mass distributions for $\tilde{t}\tilde{t}^*$ and QCD samples.}\label{fig:jetmass_comparison}
\end{figure}

We calculate the jet mass, $m_{j}$, in two ways depending on the $p_T$ of the jet. For jets with $p_T<1$ TeV, we calculate $m_j$ using the energy-momentum information from both the the tracker and the colorimeters, which is the same way as is done at the LHC. For jets with $p_T>1$ TeV, the cone size of the jet is so small that calorimeter cells in the future collider may not provide enough spatial resolution to resolve the jet constituents. Therefore, we will use the method described in Ref.~\cite{Schaetzel:2013vka, Larkoski:2015yqa}, i.e. using only the tracker energy-momentum information to calculate $m_j^{(\text{track})}$. Then the jet mass is rescaled to remove the tracker's bias for charged particles, 
\begin{equation}
m_j = m_j^{(\text{track})}\frac{p_T^{(\text{track+calorimeter})}}{p_T^{(\text{track})}}\,.
\end{equation}

In fig.~\ref{fig:jetmass_comparison}, we present the jet mass distributions for boosted top candidate jets with $p_T>1$ TeV in $\tilde{t}\tilde{t}^*$ and QCD light flavor samples. Leptonically decaying top candidate jets characterized by the presence of a hard muon ($p_T > 200$ GeV) inside the jet are shown in the top panel while hadronically decaying top candidate jets are shown in the bottom panel. In the $\tilde{t}\tilde{t}^*$ sample, the leading top candidate jets are likely from boosted top quarks produced from stop decays. This is reflected in the jet mass distribution of the leading jet in $\tilde{t}\tilde{t}^*$ which peak at $\sim m_{\text{top}}$ while QCD jets peak at much lower values as shown in fig.~\ref{fig:jetmass_comparison} (b). A similar trend is observed for leptonically decaying top candidates, as shown in fig.~\ref{fig:jetmass_comparison} (a), with a minor difference that the jet mass distribution peaks at slightly lower values $\sim 145$ GeV due to missing energy.

%%%%%%%%%%%%%%%%%%%%%%%%%%%%%%%%%%%%%%%%%%%%%%%%%%%%
\subsection{$N$-subjettiness}\label{sec:Nsubjet}
%%%%%%%%%%%%%%%%%%%%%%%%%%%%%%%%%%%%%%%%%%%%%%%%%%%%

\begin{figure}
\centering
\includegraphics[width=0.5\textwidth]{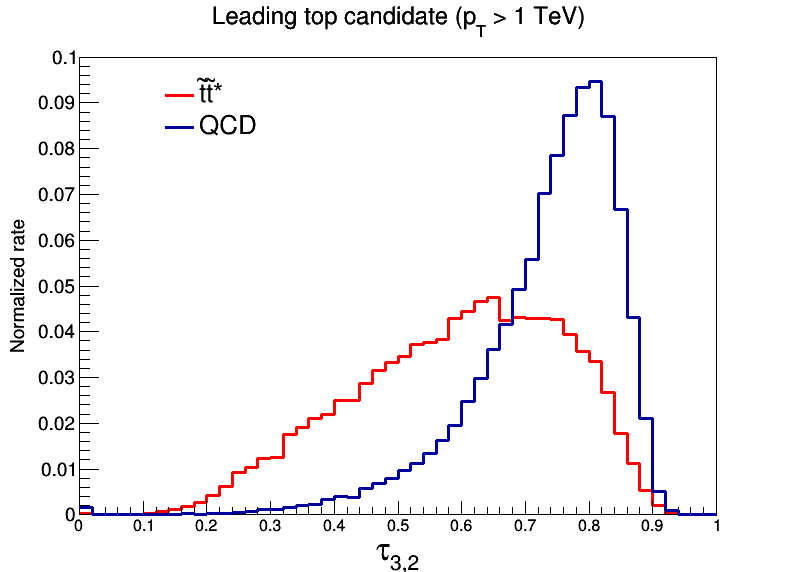}
\caption{$\tau_{3,2}$ distribution of leading top candidate jet in $\tilde{t} \tilde{t}^*$ and QCD samples. }\label{fig:tau32}
\end{figure}

A boosted top quark decaying hadronically has a three-prong substructure unlike a QCD jet. One of the jet observables that exploits this $N$-prong substructure of 
boosted particles is $N$-subjettiness $\tau_N^{(\beta)}$ which is defined as in Ref. \cite{Thaler:2010tr} : 
\begin{equation}
\tau_N^{(\beta)} = \sum\limits_{i } p_{T,i} \mathrm{min} \left\{ (\Delta R_{1,i})^{\beta}, (\Delta R_{2,i})^{\beta}, \cdots ,(\Delta R_{N,i})^{\beta} \right\} \label{eq:tau32}
\end{equation}
where the sum runs over all the constituent particles of the jet, $p_{T,i}$ is the $p_T$ of the $i^{\text{th}}$ constituent particle, $\Delta R_{J,i}$ is the angular 
separation\footnote{Angular separation is defined as $\Delta R_{J,i} = \sqrt{(\Delta \eta_{J,i})^2 + (\Delta \phi_{J,i})^2}$.} between the 
$i^{\text{th}}$ constituent and subjet axis $J$ and the $\beta$ parameter is an angular weighting exponent. The $N$ subjet axes are defined using the exclusive $k_T$-algorithm with 
WTA recombination scheme. For the case of top quark which has a 3-prong decay,  $\tau_3^{(\beta)}$ is the relevant observable. However, it has been shown in Ref.~\cite{Thaler:2010tr}
that following variable is a better discriminator between top jets and QCD jets:
\begin{equation}
\tau_{3,2}^{(\beta)} = \frac{\tau_3^{(\beta)}}{\tau_2^{(\beta)}}
\end{equation}
From here on, we will set $\beta=1$. For top candidate jets with $p_T<1$ TeV, we use both tracker and calorimeter information to compute $\tau_{3,2}$ while for jets with $p_T > 1$ TeV, we only use tracker information. In fig.~\ref{fig:tau32}, the $\tau_{3,2}$ distribution of the leading top candidate jet is shown for  $\tilde{t} \tilde{t}^*$  and QCD samples. 
For the figure, only boosted top candidate jets with $p_T>1$ TeV are selected. The $\tilde{t} \tilde{t}^*$ sample was generated for $\tilde{t}-\tilde{g}-\tilde{\chi}_1^0$ simplified model with $m_{\tilde{t}}=4$ TeV, $m_{\tilde{g}}=2$ TeV and $m_{\tilde{\chi}_1^0} = 200$ GeV. Due to 3-prong substructure of top quark decays, $\tau_{3,2}$ for boosted tops peaks at smaller values compared to that for QCD jets.

%%%%%%%%%%%%%%%%%%%%%%%%%%%%%%%%%%%%%%%%%%%%%%%%
\subsection{Mass-drop}\label{sec:massdrop}
%%%%%%%%%%%%%%%%%%%%%%%%%%%%%%%%%%%%%%%%%%%%%%

For boosted jets containing a hard muon $p_T > 200$ GeV,  mass-drop $x_\mu$ is defined as follows \cite{Thaler:2008ju, Rehermann:2010vq}: 
\begin{equation}\label{eq:massdrop}
x_\mu \equiv 1-\frac{m_{j\slashed{\mu}}^2}{m_{j}^2}\,,
\end{equation}
where $m_{j}$ is the jet mass calculated as in Sec.~\ref{sec:jetmass} and $m_{j\slashed{\mu}}$ is the mass of the jet excluding the hard muon. The observable measures how much of the jet invariant mass is carried by hadronic activity. In a boosted top jet with $W$ decaying to a muon, $m_{j\slashed{\mu}}$ is approximately the invariant mass of the $b$ jet, which is only a small fraction of $m_j \sim m_{\rm top}$. Thus we expect $x_\mu$ to be close to 1. On the other hand, for heavy flavor jets such as $b$ jets, the muon only carries a small fraction of energy and a large jet invariant mass should come from hadronic activity, resulting in $x_\mu \to 0$.
 
\onecolumngrid

\begin{figure}\centering
\centering
\includegraphics[width=0.76\textwidth]{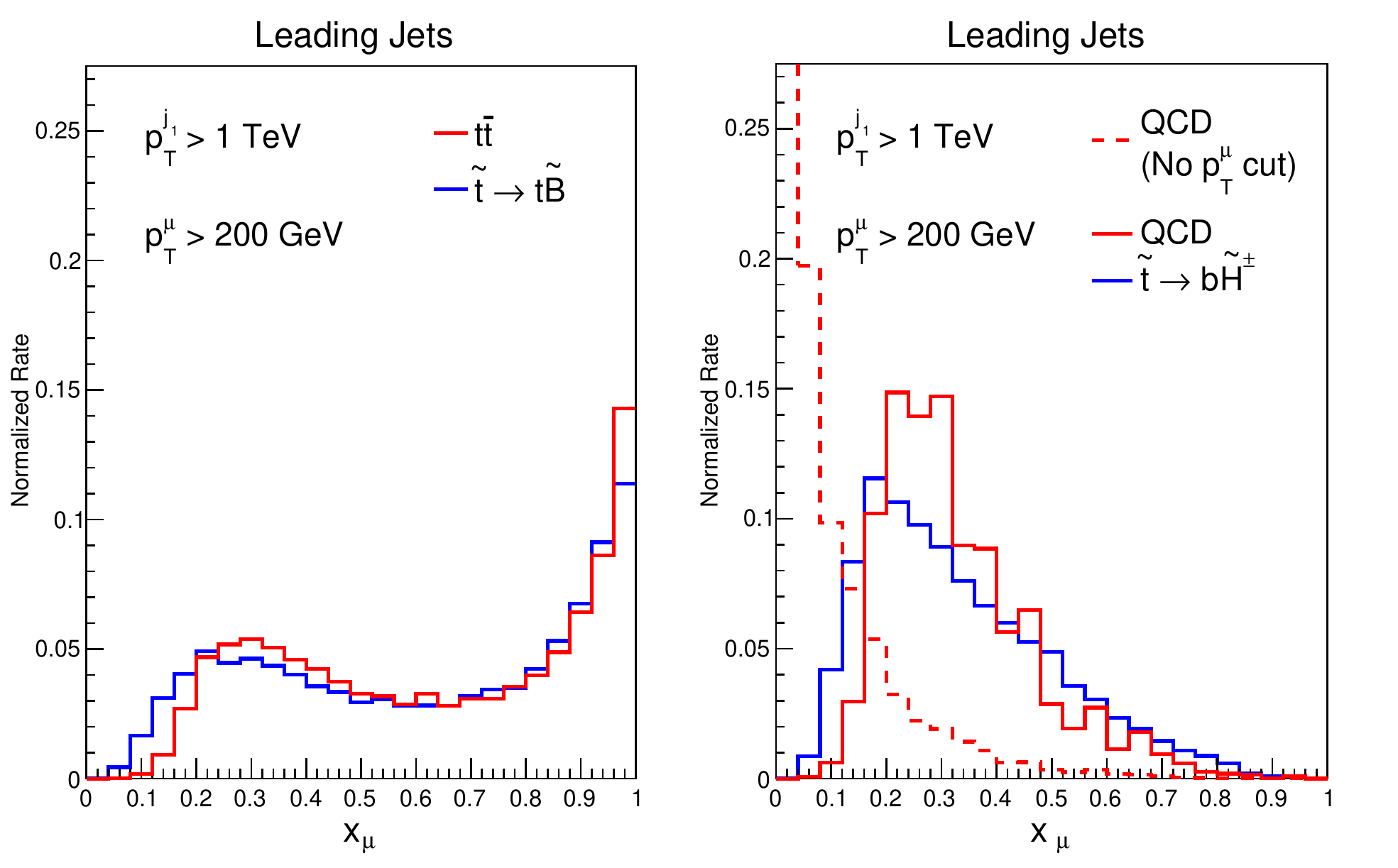}
\caption{Mass-drop distributions for the leading jet with $p_T > 1$ TeV in signals and backgrounds. Left: mass drop distributions of $t\bar{t}$ background and $\tilde{t} \to t \tilde{B}$ SUSY sample. Right: mass drop distributions of QCD background and $\tilde{t} \to b \tilde{H}^\pm$ SUSY sample. We also require the muon to have $p_T > 200$ GeV except for the dashed QCD distribution, which is obtained without any muon $p_T$ cut.  }\label{fig:massdrop}
\end{figure}

\twocolumngrid

The distributions of $x_\mu$ for different samples are presented in fig.~\ref{fig:massdrop}. The left panel shows the $x_\mu$ distributions of leading jets in the SM $t\bar{t}$ and SUSY $\tilde{t} \to t \tilde{B}$ samples. The distributions are similar and both peak at $x_\mu \approx 1$ since most of the leading jets in both samples are top jets. There is a smaller bump at lower $x_\mu$, which comes from tops with leptonic-decaying $b$'s\footnote{Leptonic decaying $b$ is defined as a $b$ jet with a muon in it.}  and hadronic decaying $W$ bosons. 
The right panel shows the distributions of leading jets in the SM QCD and SUSY $\tilde{t} \to b \tilde{H}^\pm$ samples. In these two samples, the leading jets  are mostly $b$ jets. Thus their distributions are comparable and both peak at smaller values of $x_\mu$ close to 0. Notice that the non-zero peak value of $x_\mu$ is due to the requirement that muon inside the jet satisfy $p_T>200$ GeV. The mass drop of the leading $b$ jets peaks at zero when the muon $p_T$ cut is removed, consistent with the results in Ref.~\cite{Thaler:2008ju, Rehermann:2010vq}.

%%%%%%%%%%%%%%%%%%%%%%%%%%%%%%%%%%%%%%%%%%%%%%
\section{Analysis: $\tilde{t}-\tilde{H}$ Simplified Model}
\label{sec:analysis1}
%%%%%%%%%%%%%%%%%%%%%%%%%%%%%%%%%%%%%%%%%%%%%%
In this and next sections, we will present analyses and results for the two stop simplified models. 
NLO+NNLL cross-sections were used for stop-pair and gluino-pair processes \cite{Borschensky:2014cia} while LO cross-sections from {\tt MadGraph} were used for the remaining processes\footnote{We follow Ref.~\cite{Dawson:2014pea} and treat tops as final state particles instead of using top parton distribution function in evaluating SUSY production associated with tops such as $\tilde{t} \tilde{g}$ associated production.}.

%%%%%%%%%%%%%%%%%%%%%%%%%%%%%%%%%%%%%%%%%%
\subsection{Boosted Top and Bottom Tagging}
\label{sec:combine_massdrop_jetmass}
%%%%%%%%%%%%%%%%%%%%%%%%%%%%%%%%%%%%%%%%%%

In analyzing the $\tilde{t}-\tilde{H}$ simplified model and distinguishing it from the $\tilde{t}-\tilde{B}$ model, we will combine jet mass $m_j$ and mass drop $x_\mu$ variables to tag boosted top and bottom jets. 
We define a \emph{boosted top} jet as a jet (clustered using the method in Sec.~\ref{sec:jet_finding} with $p_T > 500$ GeV) with a $p_T>200$ GeV muon inside and $x_\mu > 0.5$ {\bf or} $m_j > 120$ GeV. A \emph{boosted $b$} jet, on the other hand, is required to have a $p_T>200$ GeV muon inside and satisfy $x_\mu < 0.5$ {\bf and} $m_j < 120$ GeV. The tagging efficiencies for both the SUSY and the SM samples are shown in fig.~\ref{fig:tagging_efficiency}. 

In both taggings, the muon-in-the-jet requirement is because the decay of either a boosted bottom or top could give a hard muon close to the hadronic jet axis with a certain branching fraction. This is the same strategy as in Ref.~\cite{Cohen:2013xda}. Yet to further distinguish between $\tilde{t}-\tilde{H}$ and $\tilde{t}-\tilde{B}$ simplified models, we need to tell apart a boosted $b$ and a top jet using a combination of $m_j$ and $x_\mu$ observables. Tops with leptonic $W$'s are likely to have $x_\mu$ close to 1 but smaller $m_j$ while tops with hadronic $W$'s but leptonic $b$'s have smaller $x_\mu$ but larger $m_j$. To tag both cases and keep most of the SUSY signals after kinematic cuts, we require top jets to satisfy either $x_\mu > 0.5$ or $m_j > 120$ GeV. On the other hand, a $b$ jet has a small jet mass as well as mass drop. Thus a tagged $b$ jet is required to have $x_\mu < 0.5$ and $m_j < 120$ GeV simultaneously.

The efficiency to tag a top quark produced in the SUSY decay process $\tilde{t} \to t \tilde{\chi}^0$ ($\tilde{\chi}^0$ could be either a bino or a neutral higgsino) is around 10\%. The efficiencies of SM background events  containing top pairs is at around $1\% -3\%$. The $t \bar{t}$ sample has a smaller efficiency than the stop-pair sample because the leading jet in the top-pair sample is occasionally ISR. 
On the other hand, QCD jets are mistagged as top jets at a rate of at most  $\sim 0.1\%$ with the mistag rate even lower in the low $p_T$ bins. The high background suppression is achieved due to a hard muon-in-jet requirement. $b$ jets from $\tilde{t} \to b \tilde{H^\pm}$ are mistagged as top jets at a mere percentage level.

For boosted $b$ tagging, the efficiency to tag a $b$ jet from $\tilde{t} \to b \tilde{H}^\pm$ is around $4\% -5\%$. The SM backgrounds with tops in the final state are suppressed with efficiencies $\lesssim 0.5 \%$. The efficiency of the QCD background is even smaller at $\sim 0.2\%$. Top jets from $\tilde{t} \to t \tilde{H}^0$/$\tilde{B}$ are mistagged as bottom jets at $1\%$ level, similar to the mistag rate of $b$ jets using the top tagging strategy.

%%%%%%%%%%%%%%%%%%%%%%%%%%%%%%%%%%%%%%%%%%%%%%
\subsection{Event Selection}
\label{sec:cutflow}
%%%%%%%%%%%%%%%%%%%%%%%%%%%%%%%%%%%%%%%%%%%%%%

We require the events to satisfy the following requirements: 
\begin{itemize}
\item At least two $R=0.5$ anti-$k_T$ jets each with $p_T > 1$ TeV; 
\item No isolated lepton with $p_T > 35$ GeV. 
\item $|\Delta \phi (j, \MET)| > 1.0$ for any anti-$k_T$ jet with $p_T > 500$ GeV;
\item $\slashed{E}_T > 3.0$ TeV;
%\item A hard muon ($p_T > 200$ GeV) in any of the first 4 leading jets;
\item At least one top-tagged or bottom-tagged jet with the tagging described in Sec.~\ref{sec:combine_massdrop_jetmass}.
\end{itemize}
The lepton isolation criteria is that the total sum of $p_T$ of all the charged particles inside a cone with $R=0.5$ around the lepton is less than 10\% of the lepton's $p_T$.

\begin{figure}\centering
\centering
\includegraphics[width=0.9\textwidth]{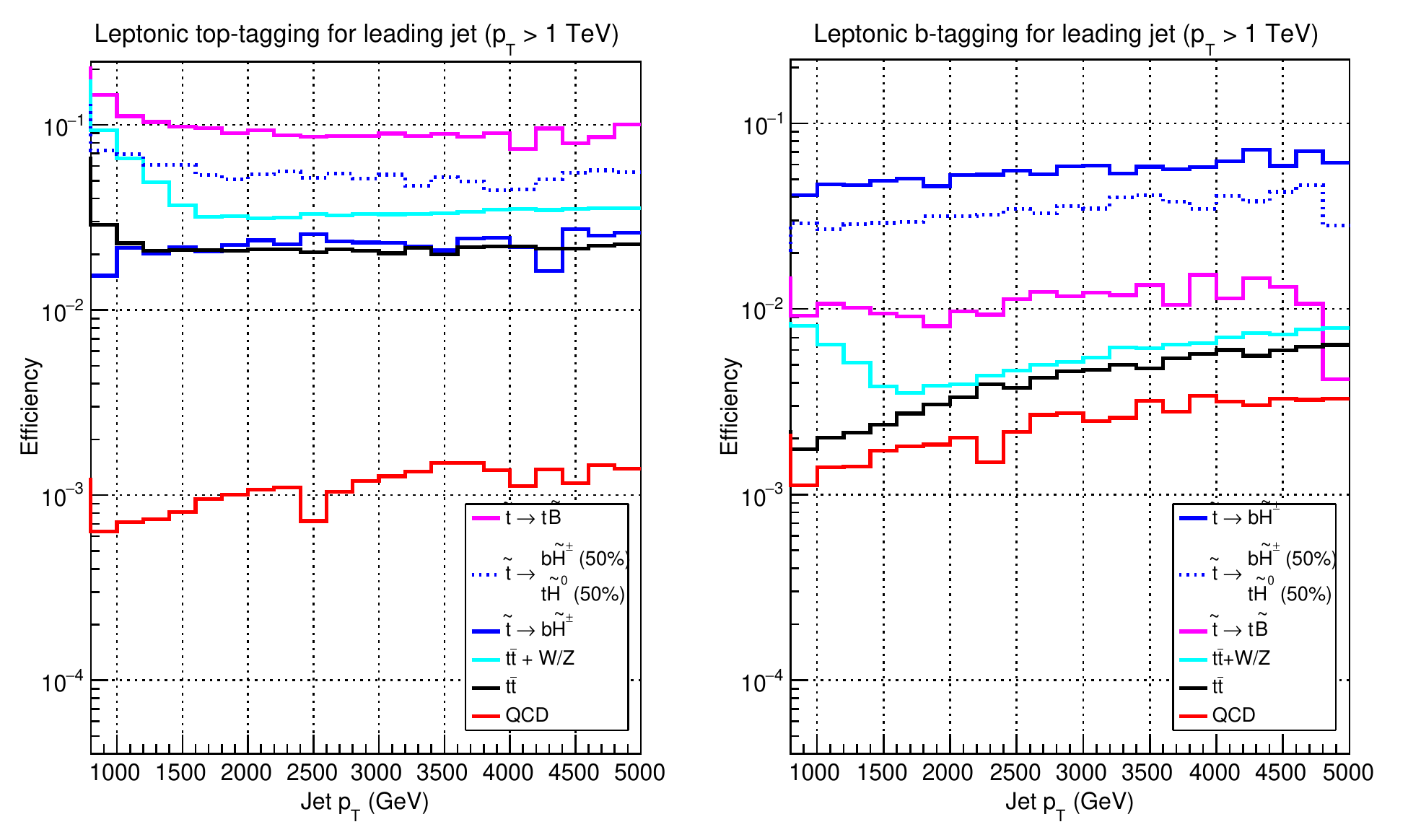}
\caption{Boosted top and $b$ tagging efficiencies for the leading jet (fraction of total events with the leading jet tagged) in different event samples as a function of jet $p_T$: SUSY events with stops decaying only to $b\tilde{H}^{\pm}$ (blue) or $t\tilde{B} $ (pink), SUSY events with stops decaying to either neutral or charged $\tilde{H}$ in the full $\tilde{t}-\tilde{H}$ simplified model (blue dotted), SM $t\bar{t}$ background (black), SM QCD background (red) and SM $t\bar{t}+W/Z$ background (light blue). We assumed $m_{\tilde{t}} = 4$ TeV and $m_{\tilde{H}}$, $m_{\tilde{B}} = 500$ GeV. }
\label{fig:tagging_efficiency}
\end{figure}

For the $\tilde{t}-\tilde{B}$ simplified model, $\tilde{t}$ only decays to $t\tilde{B}$. Given the efficiencies shown in fig.~\ref{fig:tagging_efficiency}, we expect 10\% of the signal events to be top-tagged and a negligible fraction of the events to be $b$-tagged. On the other hand, in the $\tilde{t}-\tilde{H}$ simplified model, $\tilde{t}$ decays to both $t\tilde{H}^0$ and $b\tilde{H}^\pm$, each with 50\% branching fraction. A SUSY signal event could contain either pure decays where both $\tilde{t}$'s decay though the same channel or mixed decays where one $\tilde{t}$ decays to $t\tilde{H}^0$ with the other one to $b\tilde{H}^\pm$. Since the signal efficiencies for tagging a boosted jet are $\lesssim 10\%$, the chance of tagging both $t$'s or $b$'s in the pure decay case or tagging both $t$ and $b$ in the mixed decay case is very low (typically less than 1 event after all the kinematic cuts for 3 ab$^{-1}$ of data). The signal events are then a mixture with some events $t$-tagged and the rest $b$-tagged. We will use the number of $t$ and $b$-tagged events to pin down the identity of LSP and differentiate the two simplified models in the next section.

\onecolumngrid

\twocolumngrid

\onecolumngrid

\begin{figure}
\begin{subfigure}[h]{0.42\textwidth}
\centering
\includegraphics[width=0.96\textwidth]{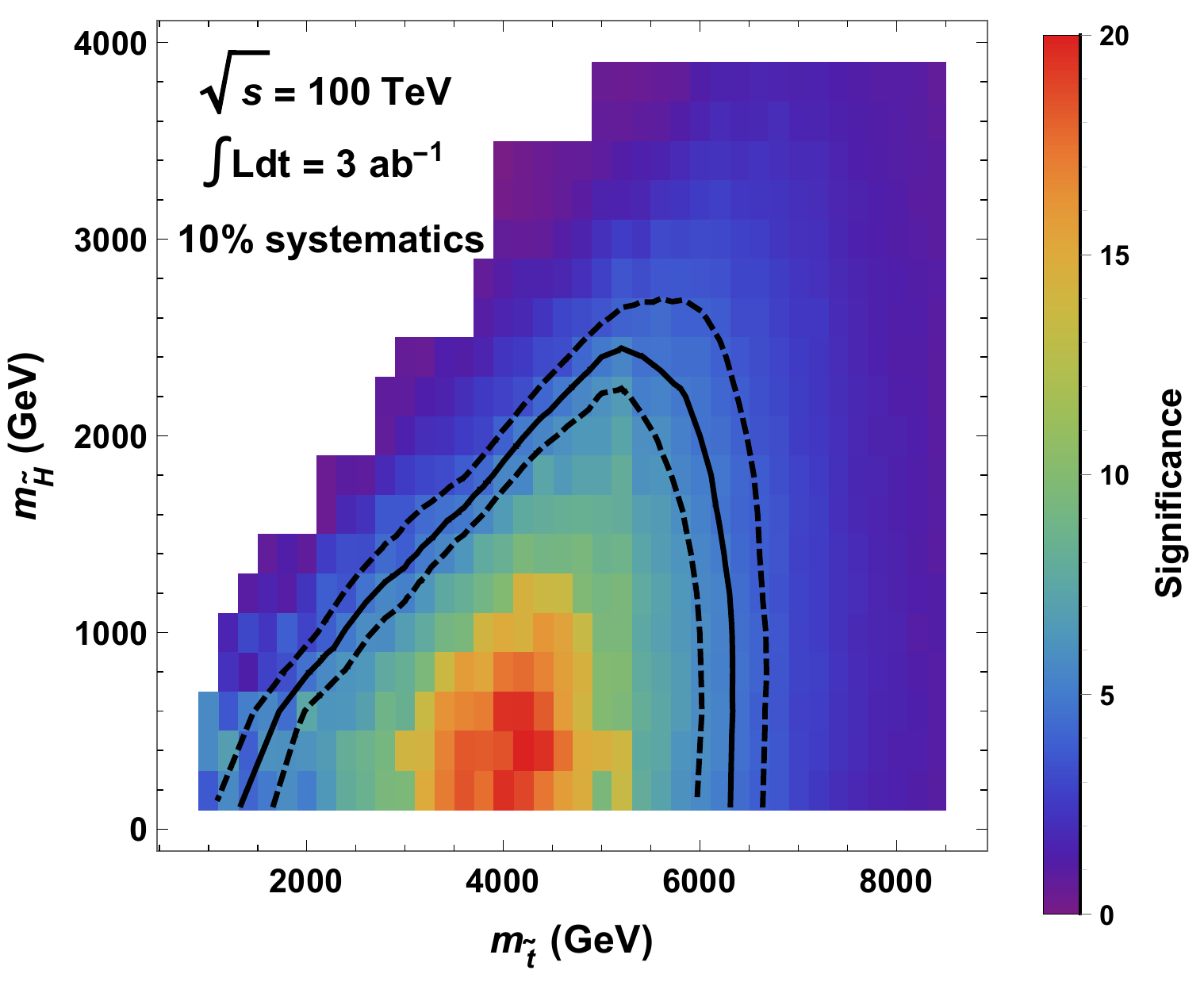}
\caption{Discovery}
\label{fig:higgsino_discovery}
\end{subfigure}
\begin{subfigure}[h]{0.42\textwidth}
\begin{center}
\includegraphics[width=1.0\textwidth]{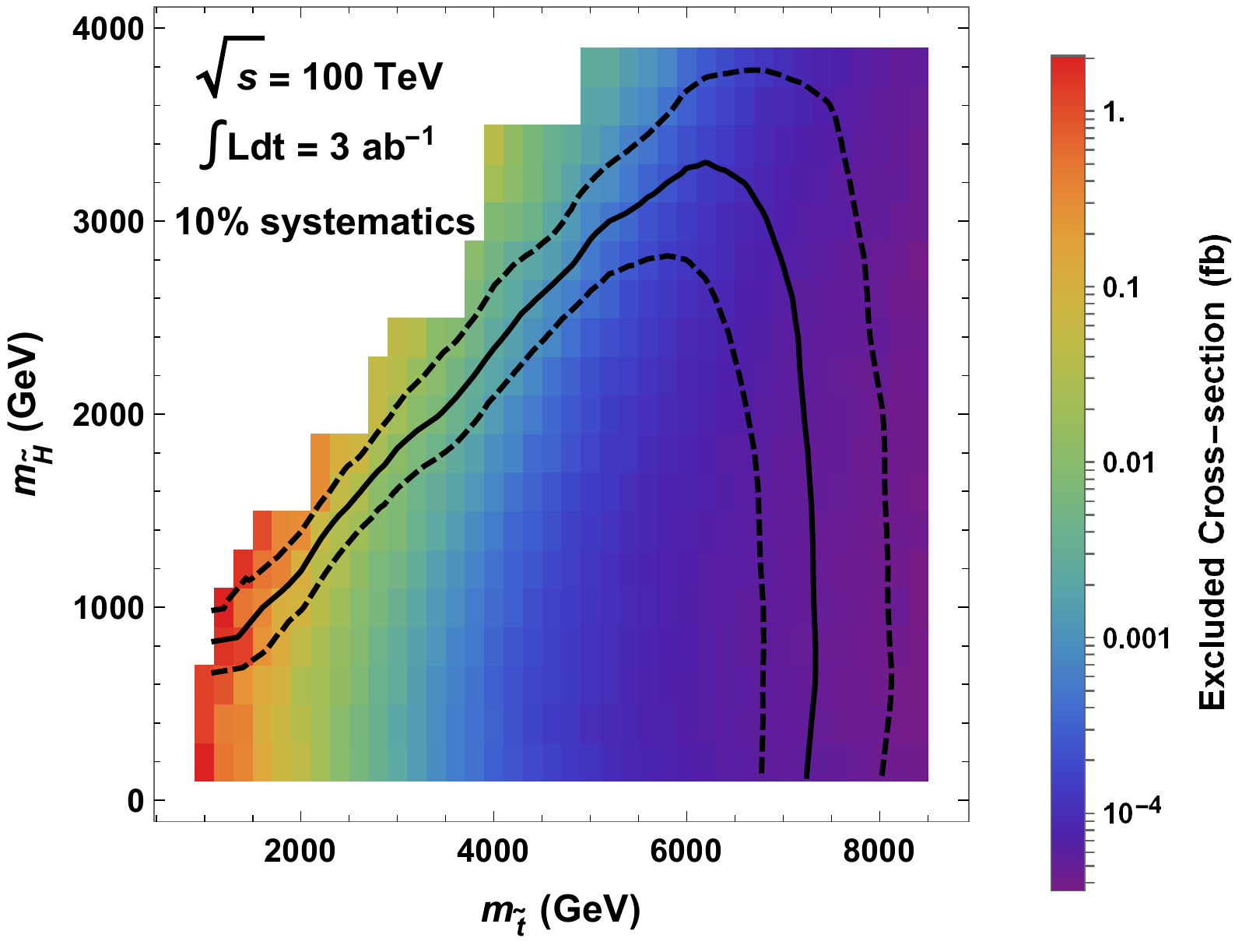}
\caption{Exclusion}
\label{fig:higgsino_exclusion}
\end{center}
\end{subfigure}
\caption{The discovery and exclusion contours for the $\tilde{t}-\tilde{H}$ simplified model at a 100 TeV collider with an integrated luminosity of 3 ab$^{-1}$. We assume a 10\% systematic uncertainty for both the signal and the background. The solid lines are 5$\,\sigma$ discovery contour (left) and exclusion at 95\% C.L.(right). The dashed lines are the $\pm 1\,\sigma$ boundaries.}
\end{figure}

\twocolumngrid
%%%%%%%%%%%%%%%%%%%%%%%%%%%%%%%%%%%%%%%%%%%%%%
\subsection{Exclusion and Discovery}
%%%%%%%%%%%%%%%%%%%%%%%%%%%%%%%%%%%%%%%%%%%%%%

We use $N_b$ to denote the number of $b$-tagged signal events after all the cuts and $N_t$ for number of $t$-tagged signal events. The total number of signal events used to set the reach is then $N_+ = N_b+N_t$. 
We scan the $(m_{\tilde{t}}, m_{\tilde{H}})$ plane and apply CL$_s$ statistics~\cite{Read:2002hq} to compute exclusion and discovery contours. Both the signal and backgrounds are modelled by Poisson statistics. A point in the mass plane is excluded if its CL$_s < 0.05$. A point could be discovered when the background only hypothesis is rejected with a $p$-value less than $3\times 10^{-7}$. We also require at least 10 total signal events for a point to be excluded or discovered. 
This conservative requirement does not affect our results when using the CL$_s$ method. It will make the physics reach estimate more robust when using the simpler approximate $S/\sqrt{B}$ estimate in the analysis in Sec.~\ref{sec:analysis2}.

From fig.~\ref{fig:higgsino_discovery}, stops with mass up to 5 - 6 TeV could be discovered for higgsino mass up to 2 TeV, assuming an integrated luminosity 3 ab$^{-1}$. At 95\% C.L., stops with mass up to 7 TeV could be excluded. All the results shown here are based on the simple cut-flows in Sec.~\ref{sec:cutflow}. We expect further optimization (e.g., through boosted decision tree) can improve the results. In addition, we do not try to perform a dedicated analysis for the compressed region where the stop mass gets closer to the higgsino mass. 

\begin{figure}\vspace{0.4cm}
\centering
\includegraphics[width=0.46\textwidth]{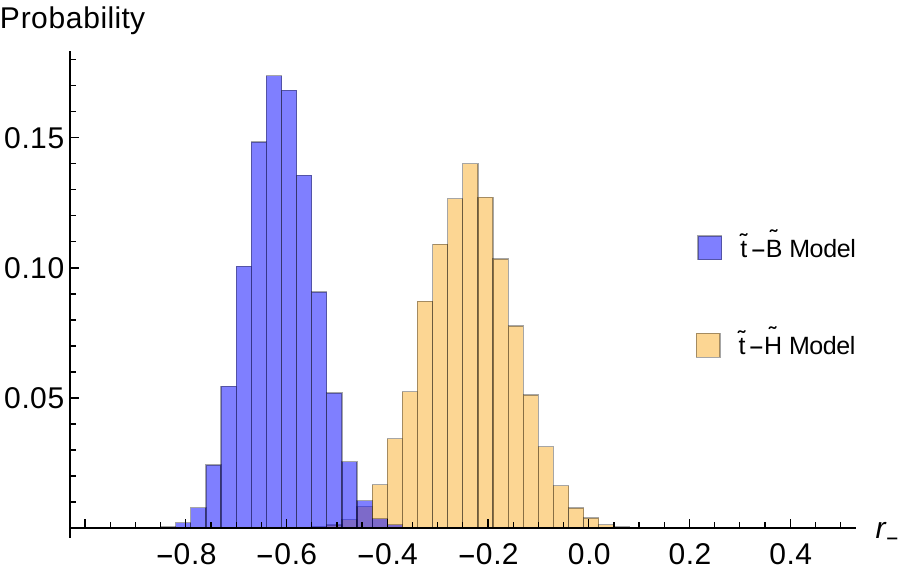}
\caption{Distributions of $r_-$ for stop-bino and stop-higgsino simplified models given $m_{\tilde{t}} = 4$ TeV, $m_{\tilde{H}}, m_{\tilde{B}} = 500$ GeV.}
\label{fig:NMinus_distribute}
\end{figure}

The total number of signal events, $N_+$, will be the same for $\tilde{t}-\tilde{H}$ and $\tilde{t}-\tilde{B}$ simplified models if the $\tilde{H}$ and $\tilde{B}$ have the same masses. Assuming a discovery of stops, we proceed to distinguish between the two simplified models using the difference between $N_b$ and $N_t$. The observable we will use is a ratio
\begin{equation}\label{eq:NMinus_def}
r_- = \frac{N_b - N_t}{N_b + N_t}\,.
\end{equation}
The advantage of a ratio observable is that systematic uncertainties contributing to individual observables are likely to cancel out. The distributions of $r_-$ for $\tilde{t}-\tilde{H}$ and $\tilde{t}-\tilde{B}$ simplified models are demonstrated in fig.~\ref{fig:NMinus_distribute}. In the figure, $r_-$ peaks at $\sim - (0.2-0.3)$ in the $\tilde{t}-\tilde{H}$ model. This can be understood as follows: the $t$-tagging efficiency of the signal is $\epsilon^{t}_{\rm sig} \approx $ 10\% while that for $b$-tagging efficiency is about $\epsilon^{b}_{\rm sig} \approx 5$\%, as in Sec.~\ref{sec:combine_massdrop_jetmass}. In a $\tilde{t}-\tilde{H}$ sample, 1/4 of the events contain two $b$ jets, 1/4 of the events contain two $t$ jets while the rest half contains one $b$ and one $t$ jet. Thus $N_b \approx \epsilon^{b}_{\rm sig} N_+$ and $N_t \approx \epsilon^{t}_{\rm sig} N_+$, leading to $r_- \approx -0.3$. In contrast, $r_-$ peaks at $\sim -0.6$ in the $\tilde{t}-\tilde{B}$ model. This is consistent with that almost all the events in a $\tilde{t}-\tilde{B}$ sample contain two boosted $t$ jets. Ignoring the rate of mistagging a $t$ jet as a $b$ jet, $r_- \approx -1$. Including the mistag rate shifts the central value to $-0.6$. 

\begin{figure}\vspace{0.3cm}
\centering
\includegraphics[width=0.42\textwidth]{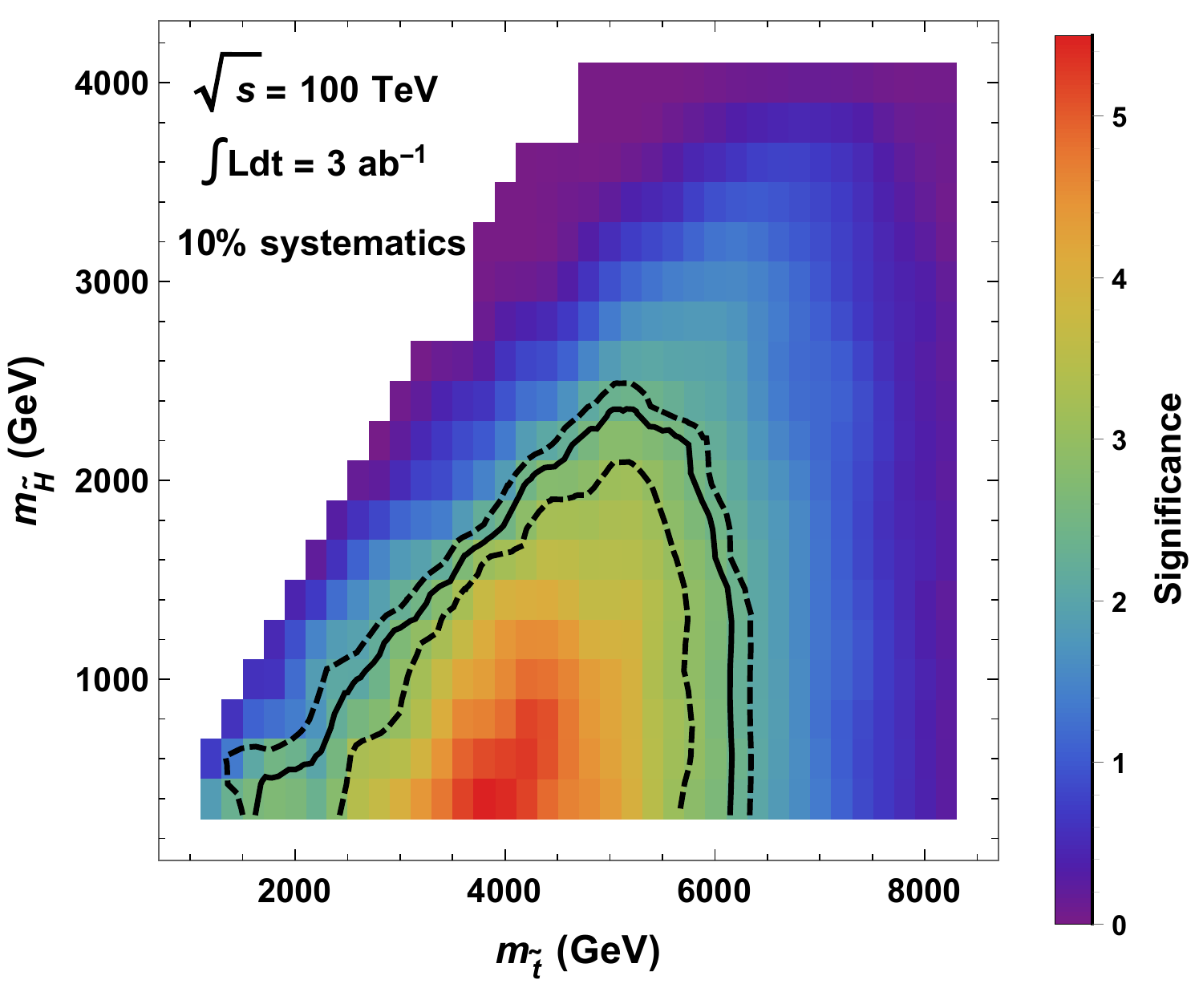}
\caption{The 2$\sigma$ exclusion contour of $\tilde{t}-\tilde{B}$ simplified model based on $r_-$ assuming the signal events are from $\tilde{t}-\tilde{H}$ model. The dashed contours are $\pm 1\,\sigma$ boundaries.}
\label{fig:model_exclusion}
\end{figure}

Finally we show the 95\% C.L. exclusion of $\tilde{t}-\tilde{B}$ model based on $r_-$ assuming that the signal comes from the $\tilde{t}-\tilde{H}$ simplified model in fig.~\ref{fig:model_exclusion}. From the figure, one could see that the 95\% C.L. contour overlaps with the $5\sigma$ discovery reach in fig.~\ref{fig:higgsino_discovery}. Thus using $N_+$ and $r_-$, we could not only discover stops up to 6 TeV but also determine whether the LSP is a higgsino or bino. 

%%%%%%%%%%%%%%%%%%%%%%%%%%%%%%%%%%%%%%%%%%%%%%
\section{Analysis: $\tilde{t}-\tilde{g}-\tilde{\chi}_1^0$ Simplified Model}
\label{sec:analysis2}
%%%%%%%%%%%%%%%%%%%%%%%%%%%%%%%%%%%%%%%%%%%%%%

\subsection{Top-tagging}
\label{sec:toptagging2}
The final state in the $\tilde{t}-\tilde{g}-\tilde{\chi}_1^0$ simplified model is characterized by 6 top quarks, several of which may be boosted. Therefore, we rely on multiple top tags to discriminate signal from background. Anti-$k_T$ jets with $R=0.5$ and C/A jets with $R=1.0$ are identified using the procedure outlined in Sec. \ref{sec:jet_finding}. 
 Two separate top tagging strategies are used for hadronic and leptonic top decays.  If an energetic muon with $p_T>200$ GeV is among the constituents of a C/A jet, that jet is identified as a leptonically decaying top candidate. Otherwise, the C/A jets are reclustered using $p_T$ dependent radius following the two-step scaling down procedure described in Section \ref{sec:jet_finding}. The resulting subjets are identified as hadronically decaying top candidates. For hadronic top candidates, jet mass is required to lie in the top mass window of $140$ GeV $<m_j<$ $240$ GeV to reject QCD jets as shown in fig \ref{fig:mass_hadronic}. For leptonic top candidates, top mass reconstruction is not 
possible due to missing energy. Nevertheless, requiring $m_j>75$ GeV provides a good discrimination between top jets and QCD jets as shown in fig. \ref{fig:mass_leptonic}.

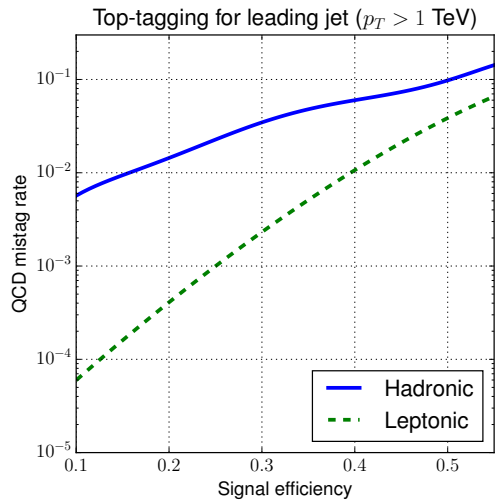
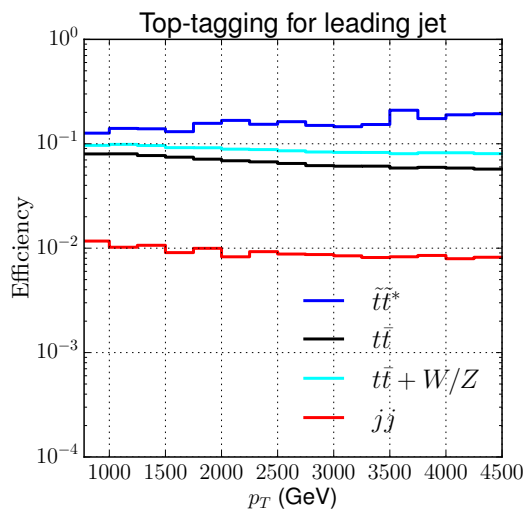
\begin{figure}
    \centering
    \begin{subfigure}[!h]{0.40\textwidth}
        \includegraphics[width=\textwidth]{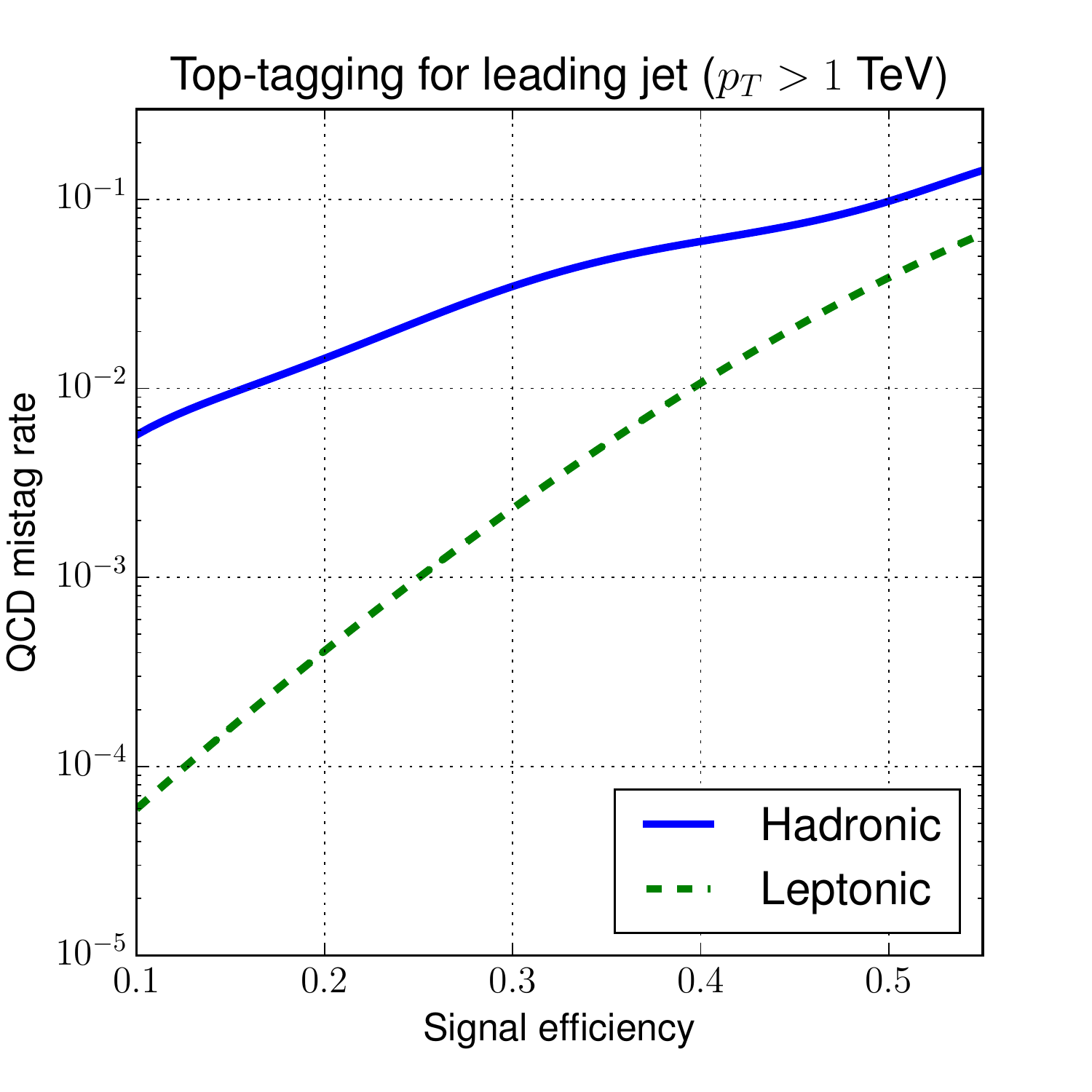}
        \caption{}
        \label{fig:mistag}
    \end{subfigure}
    ~ \qquad
    \begin{subfigure}[!h]{0.40\textwidth}
        \includegraphics[width=\textwidth]{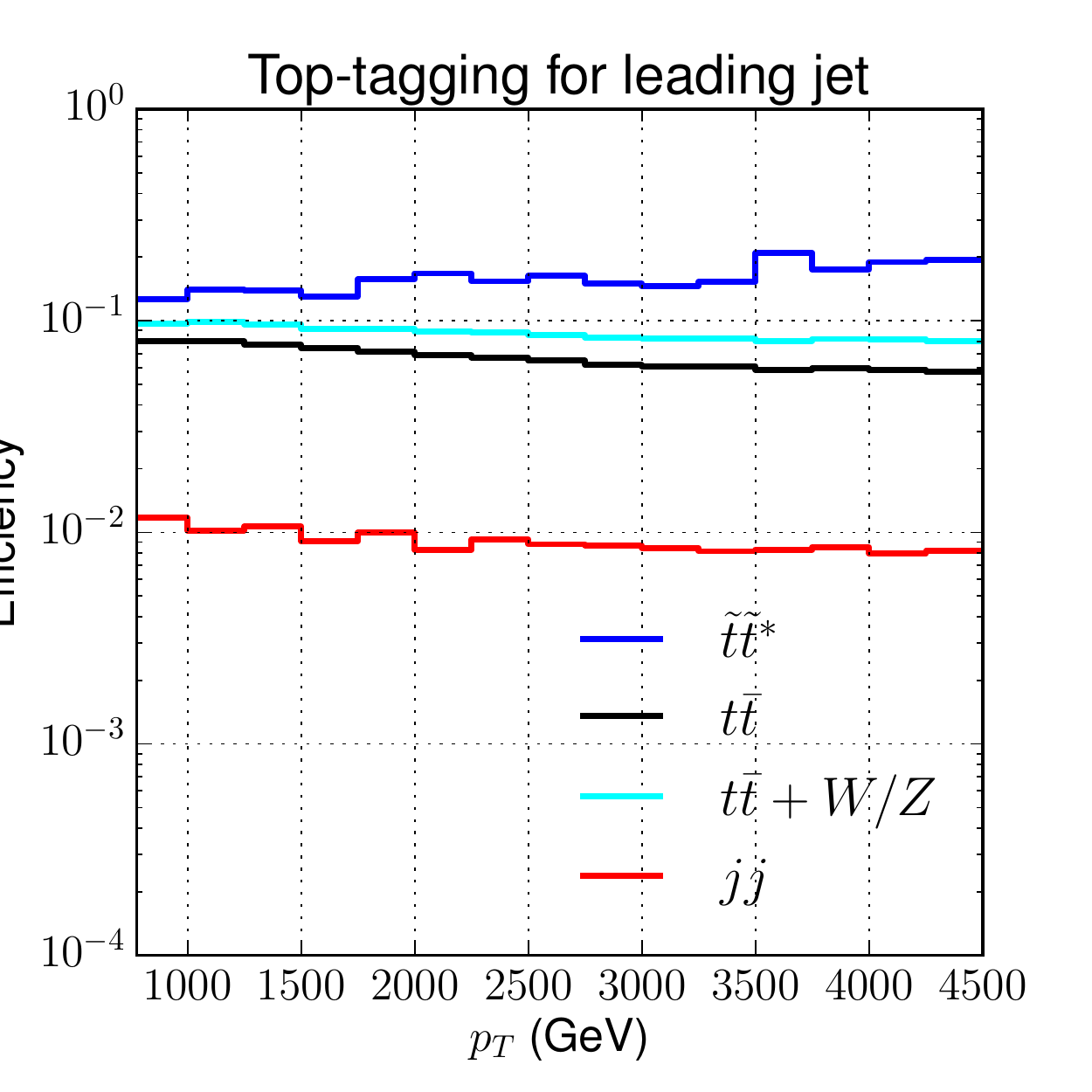}
        \caption{}
        \label{fig:top_tag}
    \end{subfigure}
            \begin{tikzpicture}[overlay]
        		\node[rotate=90,black] at (0,0){};
        		\draw [fill=white,white] (-8,-1) rectangle (-7.0,2);
			\node[rotate=90,black] at (-7.2,0.5){\text{Efficiency}};
		\end{tikzpicture}
    \caption{(a) QCD mistag rate vs signal efficiency for top-tagging the leading top candidate jet. 
    		 (b) Top-tagging efficiency for signal and SM processes as a function of jet $p_T$.}
\label{fig:top_tag_mistag}
\end{figure}

To further improve top-tagging, we use the $N$-subjettiness variable $\tau_{3,2}$ (see Section \ref{sec:Nsubjet}) for hadronic top decays and the mass drop variable $x_\mu$ (see Section \ref{sec:massdrop}) for leptonic top decays. By imposing cuts on these two parameters, it is possible to obtain the desired signal efficiency. 
In fig \ref{fig:mistag}, the QCD mistag rate is plotted against signal efficiency for the leading top candidate jet. Cuts on the jet mass for both leptonic and hadronic channels 
are already imposed and included in the efficiency and mistag rates. Note that in computing the rates in the top panel of fig.~\ref{fig:top_tag_mistag}, we used slightly different definitions from those for fig. 10 and the bottom panel of fig.~\ref{fig:top_tag_mistag}: the efficiencies of hadronic (leptonic) top tagging are the fractions of events with a hadronic (leptonic) top candidate satisfying the tagging requirements. 
While top-tagging is more efficient in the leptonic channel, it suffers from a low branching ratio.  
Therefore, using both leptonic and hadronic tagging is beneficial. We will choose $0.1 < \tau_{3,2} < 0.45$ and $x_\mu > 0.7$ which corresponds to QCD mistag rate
of $\sim 1 \%$. Using these cuts, the jet-$p_T$ dependence of the combined top-tagging efficiency for the leading top candidate is plotted in fig. \ref{fig:top_tag}. The combined 
signal efficiency is $\sim10-20 \%$ compared to $\sim 1 \%$ for QCD jets.

%%%%%%%%%%%%%%%%%%%%%%%%%%%%%%%%%%%%%%%%%%%%%%
\subsection{Event Selection and Results}
%%%%%%%%%%%%%%%%%%%%%%%%%%%%%%%%%%%%%%%%%%%%%%

\begin{table*}
\setlength{\extrarowheight}{1pt}
\setlength{\tabcolsep}{8pt}
\centering
    \begin{tabular}{ c  c  c  c  c }
    \hline\hline
    Cuts	 					 & $\tilde{t}\tilde{t}^*$ 	& $\tilde{t}\tilde{g}$		& $\tilde{g}\tilde{g}$		& $\tilde{g}\tilde{g}+t\bar{t}$ 	\\
    \hline 
    $H_T>4$ TeV, $\MET>250$ GeV	 & $8809$				& $12415$			& $8.94 \times 10^6$	& $34990$				\\
    No leptons					 & $7687$				& $10723$			& $8.08  \times 10^6$	& $30312$				\\
    $n_j\ge7$ and ISR cuts		 & $3574$				& $4435$				& $1.13  \times 10^6$	& $10517$				\\	
    $|\Delta \phi (j, \MET)| > 0.5$	 & $2788$				& $3589$				& $901151$			& $8294$					\\
    $1$ top-tag					 & $490$				& $630$				& $131816$			& $1412$					\\
    $2$ top-tags					 & $228$				& $233$				& $27910$			& $500$					\\
    $3$ top-tags					 & $52$				& $48$				& $3555$				& $111$					\\
    $H_T,\MET$ cuts				 & $8.6$				& $2.1$				& $0$				& $1.5$					\\
    \hline\hline
    \end{tabular}
\caption{Cut flow for SUSY processes at $\sqrt{s}=100$ TeV and $\LL = 30$ ab$^{-1}$.  SUSY masses are $m_{\tilde{t}}=5.5$ TeV, $m_{\tilde{g}}=2.75$ TeV and $m_{\tilde{\chi}_1^0} = 200$ GeV.}
\label{table:cutflow_SUSY}
\end{table*}

\begin{table*}
\setlength{\extrarowheight}{1pt}
\setlength{\tabcolsep}{8pt}
\centering
    \begin{tabular}{ c  c  c  c  c  c }
    \hline\hline
    Cuts	 					 & $t\bar{t}$			& $t\bar{t}+W/Z$		& QCD				& $t+W/Z$			& $W/Z$+jets 				\\ 
    \hline
    $H_T>4$ TeV, $\MET>250$ GeV	 & $5.96 \times 10^7$	& $3.94 \times 10^6$	& $1.24 \times 10^9$	& $8.32 \times 10^6$	& $8.65 \times 10^7$		\\ 
    No leptons					 & $5.72 \times 10^7$	& $3.76 \times 10^6$	& $1.24  \times 10^9$	& $8.06 \times 10^6$	& $8.34 \times 10^7$		\\ 
    $n_j\ge7$ and ISR cuts		 & $3.16 \times 10^6$	& $1.67 \times 10^5$	& $3.64  \times 10^7$	& $1.18 \times 10^5$	& $1.89 \times 10^6$		\\ 	
    $|\Delta \phi (j, \MET)| > 0.5$	 & $1.53 \times 10^6$	& $81624$			& $1.49  \times 10^7$	& $52675$			& $8.50 \times 10^5$		\\ 
    $1$ top-tag					 & $70782$			& $5698$				& $193858$			& $2522$				& $9589$					\\ 
    $2$ top-tags					 & $9520$				& $690$				& $479$				& $99$				& $701$					\\ 
    $3$ top-tags					 & $132$				& $18$				& $0$				& $0$				& $0$					\\ 
    $H_T,\MET$ cuts				 & $0$				& $0.1$				& $0$				& $0$				& $0$					\\
    \hline\hline
    \end{tabular}
\caption{Cut flow for SM processes at $\sqrt{s}=100$ TeV and $\LL = 30$ ab$^{-1}$.}  \label{table:cutflow_SM}
\end{table*}

The following cuts are used to discriminate between signal and background :
\begin{itemize}
\item $H_T>4$ TeV and $\MET>250$ GeV;
\item No isolated leptons with $p_T > 50$ GeV;
\item At least $7$ jets (anti-$k_T$ with $R=0.5$ and $p_T>200$ GeV); 
\item At most one ISR jet among the leading 6 jets (see below);
\item $|\Delta \phi (j, \MET)| > 0.5$ for the leading two jets;
\item At least 3 top tagged jets with the top tagging described in Section~\ref{sec:toptagging2};
\item Optimized $H_T$ and $\MET$ cuts (see below).
\end{itemize}
 At $100$ TeV collider, imposing jet multiplicity cut is not sufficient to distinguish hard jets from ISR. To resolve this issue, ISR jets are identified by one of the two criteria~\cite{Krohn:2011zp}:
\begin{itemize}
\item high rapidity : $|\eta| > 2$ 
\item a big hierarchy in successive jet $p_T$'s: for $p_T$-ordered jets, every ratio of successive jet $p_T$s less than $0.2$ is counted as an ISR.  
\end{itemize}
In the last step, harder $H_T$ and $\MET$ cuts are imposed and optimized so as to maximize the reach $\sigma$ defined as :
\begin{equation}
\sigma = \frac{S}{\sqrt{B+\gamma^2(S^2+B^2)}}
\end{equation}
where $S$ is the number of signal event, $B$ is the number of background events and $\gamma$ is the systematic uncertainty for both signal and background.

The cut flow for SUSY and SM processes at $\sqrt{s}=100$ TeV and luminosity $\LL=30 \, \text{ab}^{-1}$ is shown in Tables \ref{table:cutflow_SUSY} and  \ref{table:cutflow_SM} respectively. 
The SUSY mass spectrum is chosen to be $m_{\tilde{t}}=5.5$ TeV, $m_{\tilde{g}}=2.75$ TeV and $m_{\tilde{\chi}_1^0} = 200$ GeV.
Preliminary $H_T$ and $\MET$ cuts are designed to suppress SM backgrounds which have very large cross-sections. The signal processes  $\tilde{t}\tilde{t}^*$ and  $\tilde{t}\tilde{g}$
have up to $6$ top quarks in the final state while the SM backgrounds and the $\tilde{g}\tilde{g}$ background have fewer hard partons in the final state. This justifies the requirement for $7$ hard jets.
Nevertheless, the preliminary $H_T$ cut inadvertently selects background events with ISR jets which can mimic hard jets. Therefore, the hard jet-multiplicity cut has to be supplemented by vetoing 
ISR jets. To this end, we require that at most one ISR jet be present among the $6$ hardest jets. The $|\Delta \phi (j, \MET)|$ cut is designed to suppress SM backgrounds such as $W/Z+$ jets where the missing energy from $W/Z$ decay is mostly aligned with the leading jets due to collinear emission of $W/Z$ bosons from jets. 
In addition, it could suppress QCD mismeasurement backgrounds.

At this stage, several background processes still have $3$ orders of magnitude 
more events than the signal with QCD being the dominant background. Next, we make use of the high top-quark multiplicity in the signal processes unlike SM backgrounds that have at  
most $2$ top quarks in the final state. By requiring 3 top tags, the largest QCD background is completely eliminated while also suppressing other backgrounds.  After top-tagging, the dominant background is  $\tilde{g}\tilde{g}$ along with sub-dominant contributions from  $t\bar{t}$, $\tilde{g}\tilde{g}+t\bar{t}$  and $t\bar{t}+W/Z$ processes. In the last step, we maximize the stop reach significance by performing a scan over $H_T$-$\MET$ cuts.

The stop reach for $\tilde{t}-\tilde{g}-\tilde{\chi}_1^0$ simplified model at $\sqrt{s}=100$ TeV and luminosity of $30$ ab$^{-1}$ is shown in Table \ref{table:reach}. The NLL+NLO gluino-pair cross-section is $1.33$ pb for $2.75$ TeV gluinos at $\sqrt{s}=100$ TeV while stop-pair cross-sections are shown in Table \ref{table:reach}. For $m_{\tilde{t}}=5.5 \,(6.0)$ TeV and $m_{\tilde{g}}=2.75$ TeV, we were able to obtain a reach of $6.3 \,(3.5) \, \sigma$ for a systematic uncertainty $\gamma=0.1$. The optimal $H_T$-$\MET$ cuts were found to be $H_T>9.5$ TeV and $\MET\gtrsim1.5$ TeV (1.25 TeV for $m_{\tilde{t}}=6.0$ TeV).

\begin{table}[h]
\setlength{\extrarowheight}{1pt}
\setlength{\tabcolsep}{6pt}
\centering
	\begin{tabular}{ c  c  c  c  c }
	\hline\hline
	$m_{\tilde{t}}$ (TeV) 	& $\sigma_{pp \rightarrow \tilde{t}\tilde{t}^*}^{\text{NLO+NLL}}$ (fb)	& $S$		& $B$			& $\sigma$	\\ \hline
	$5.5$				& $0.40$						& $10.7$		& $1.7$			& $6.3$		\\
	$6.0$				& $0.23$						& $10.0$		& $6.7$			& $3.5$		\\ \hline\hline
	\end{tabular}
\caption{Stop reach for $\tilde{t}-\tilde{g}-\tilde{\chi}_1^0$ simplified model at $\sqrt{s}=100$ TeV with luminosity $\LL = 30$ ab$^{-1}$ and 
	 systematic uncertainty $\gamma=0.10$. Here, $m_{\tilde{g}}=2.75$ TeV and $m_{\tilde{\chi}_1^0} = 200$ GeV.} \label{table:reach}

\end{table}

\onecolumngrid

\begin{table}[h]
\centering
\setlength{\extrarowheight}{1pt}
\setlength{\tabcolsep}{8pt}
    \begin{tabular}{ c  c  c  c  c  c  c }
    \hline\hline
    Cuts	 					 &$\tilde{g}\tilde{g}$		& $t\bar{t}$			& $t\bar{t}+W/Z$		& QCD				& $t+W/Z$			&$W/Z$+jets 				\\ \hline

    $H_T>4$ TeV, $\MET>250$ GeV	 &$802$				& $5.96 \times 10^6$	& $3.94 \times 10^5$	& $1.24  \times 10^8$	& $8.32  \times 10^5$	& $8.65 \times 10^6$		\\ 
    No leptons					 &$764$				& $5.72 \times 10^6$	& $3.76 \times 10^5$	& $1.24  \times 10^8$	& $8.06  \times 10^5$ 	& $8.34 \times 10^6$		\\
    $n_j\ge5$ and ISR cuts	 	 &$528$				& $2.19 \times 10^6$	& $1.38 \times 10^5$	& $3.13  \times 10^7$	& $1.00  \times 10^5$	& $2.02 \times 10^6$		\\ 	
    $|\Delta \phi (j, \MET)| > 0.5$	 &$447$				& $8.97 \times 10^5$	& $57806$			& $9.74  \times 10^6$	& $38576$			& $7.69 \times 10^5$		\\ 
    $1$ top-tag					 &$88$				& $49343$			& $4804$				& $87361$			& $1951$				& $10789$				\\ 
    $2$ top-tags					 &$34$				& $5342$				& $632$				& $1352$				& $98$				& $351$					\\
    $H_T,\MET$ cuts				 &$12.4$				& $0.57$				& $0.23$				& $0$				& $0$				& $0$					\\
    \hline\hline
    \end{tabular}
\caption{Cut flow for gluino-pair and SM processes at $\sqrt{s}=100$ TeV and $\LL = 3$ ab$^{-1}$.  SUSY masses are $m_{\tilde{g}}=10$ TeV and $m_{\tilde{\chi}_1^0} = 200$ GeV.}  \label{table:cutflow_GlGl}
\end{table}

\twocolumngrid

%%%%%%%%%%%%%%%%%%%%%%%%%%%%%%%%%%%%%%%%%%%%%%
\subsection{Improvement in Gluino Search}
%%%%%%%%%%%%%%%%%%%%%%%%%%%%%%%%%%%%%%%%%%%%%%

\begin{table}
\setlength{\tabcolsep}{6pt}
\setlength{\extrarowheight}{1pt}
\centering
	\begin{tabular}{ c  c  c  c  c  c }
	\hline\hline
	$m_{\tilde{g}}$ (TeV) 	& $\sigma_{pp \rightarrow \tilde{g}\tilde{g}}^{\text{NLO+NLL}}$ (fb)	& Top tags	& $S$		& $B$			& $\sigma$	\\ 
	\hline $10.0$				& $0.31$			& $2$			& $12.4$		& $0.8$			& $8.1$		\\ 
	$11.0$				& $0.13$			& $1$			& $13.8$		& $9.5$			& $3.9$		\\ \hline \hline
	\end{tabular}
\caption{Gluino reach for $\tilde{t}-\tilde{g}-\tilde{\chi}_1^0$ simplified model at $\sqrt{s}=100$ TeV with luminosity $\LL = 3$ ab$^{-1}$ and 
	 systematic uncertainty $\gamma=0.10$. Here, $m_{\tilde{\chi}_1^0} = 200$ GeV while $m_{\tilde{t}} \gg m_{\tilde{g}}$.} \label{table:gluino_reach}
\end{table}
It should be noted that the jet observables presented so far can also be used to improve gluino reach at future hadron colliders. In Table \ref{table:cutflow_GlGl}, a cut flow analysis is presented for gluino-pair and SM processes at $\sqrt{s}=100$ TeV and luminosity $\LL=3 \, \text{ab}^{-1}$. The SUSY mass spectrum is chosen to be  $m_{\tilde{g}}=10$ TeV and $m_{\tilde{\chi}_1^0} = 200$ GeV. The only differences compared to the stop cut flow analysis is that the minimum number of jets requirement is relaxed to $5$, up to two ISR jets are allowed and at most two top tags are required. In Table \ref{table:gluino_reach}, the gluino reach at $100$ TeV collider and luminosity of $3$ ab$^{-1}$ is presented. The final $H_T$-$\MET$ optimized cuts were chosen to be $H_T>11$ TeV and $\MET>3$ TeV yielding a reach of $8.1 \, (3.9) \, \sigma$ for $m_{\tilde{g}}=10 \, (11)$ TeV assuming systematic uncertainty $\gamma=0.1$ for both signal and background. Two top tags are used for  $m_{\tilde{g}}=10$ TeV while only one top tag is used for  $m_{\tilde{g}}=11$ TeV. Compared to the same-sign di-lepton search in~\cite{Cohen:2013xda}, which could reach $\sim9$ TeV gluino assuming zero pile-up, our strategy could be sensitive to smaller production cross section and higher gluino mass.

%%%%%%%%%%%%%%%%%%%%%%%%%%%%%%%%%%%%%%%%%%%%%%
\section{Conclusions and Outlook}
\label{sec:con}
%%%%%%%%%%%%%%%%%%%%%%%%%%%%%%%%%%%%%%%%%%%%%%

The discovery of a 125 GeV Higgs boson at the LHC is a milestone in particle physics. Yet the absence of new physics signals at the LHC so far makes the existence of such a light scalar confusing. A new energy frontier is needed to resolve mysteries related to electroweak symmetry breaking and to obtain a more definite answer to whether the weak scale is tuned. Understanding the physics cases and search challenges at a future collider serve as first steps to construct this next-generation machine.

In this paper, we focus on reach of two stop simplified models at a future 100 TeV collider. Stops are key ingredients of low-energy SUSY and their mass scale directly tells us the degree of electroweak fine-tuning. In the first simplified model we study, stops are pair-produced and decay to top or bottom plus higgsinos. In the other model with gluino lighter than the stops, stops could be produced either in pairs or associated with a gluino. They will subsequently decay through gluinos to tops plus bino. The main new features of these simplified models are that the final states contain a lot of highly boosted top or bottom jets with $p_T$ above a TeV. To suppress the SM top backgrounds and for the second simplified model, SUSY backgrounds, we study and apply several simple jet observables such as track-based jet mass, $N$-subjettiness and mass drop. Combining these jet observables gives us effective tagging strategies for boosted tops and bottoms. We find that assuming 10\% systematic uncertainties, the future 100 TeV collider can discover (exclude) stops with masses up to 6 (7) TeV with 3 ab$^{-1}$ of integrated luminosity if the stops decay to higgsinos. In the second simplified model with light gluinos and the stops decay through gluinos, due to additional SUSY backgrounds from gluino pair production, a higher luminosity of about 30 ab$^{-1}$ is needed to discover stops up to 6 TeV. We could use jet observables to tell apart simplified models with different LSPs, for instance, $\tilde{t}-\tilde{H}$ model and $\tilde{t}-\tilde{B}$ model. In addition, the top tagging allows us to improve the gluino reach close to $11$ TeV with 3 ab$^{-1}$ data. 

This paper is the first one to apply jet substructure techniques at a 100 TeV collider to study (supersymmetric) top partners, which indicates the level of electroweak fine-tuning, one of the major physics questions that a future hadron collider hopefully can give a qualitative answer. Studies on applying jet substructure to search for other possible new particles at a 100 TeV collider could be found in Ref.~\cite{PhysRevD.92.054033, PhysRevD.92.073013, PhysRevD.93.075037, PhysRevD.94.095016}. Jet substructure techniques provide us a powerful way to discriminate intricate new physics final states containing many hyper-boosted objects from messy SM and SUSY backgrounds, for which the traditional search strategies may not work. The jet tools could also help us distinguish between different new physics models and improve their reach significantly, exploring further the power of the future energy frontier.

While we focus on the study of mass reach of stops, the jet observables we study could be applied to search for other new particles such as fermionic top partners, which suffer from similar issues from hyper-boosted SM objects. They may also be used in exploring new mechanisms at future colliders such as measuring the gluino decays to test whether the minimal supersymmetric SM explains the Higgs mass \cite{Agrawal:2017zwh}. In addition, the hyper-boosted top or bottom tagging may be further improved as discussed in Ref.~\cite{Bressler:2015uma}.

%%%%%%%%%%%%%%%%%%%%%%%%%%%%%%%%%%%%%%%%%%%%%%
\begin{acknowledgments}
%%%%%%%%%%%%%%%%%%%%%%%%%%%%%%%%%%%%%%%%%%%%%%
We thank Tim Cohen, Raffaele Tito D'Agnolo, James Hirschauer, Matt Low and John Stupak for useful correspondences. We also thank Matt Reece for reading and commenting on the manuscript. JF is supported by the DOE grant DE-SC-0010010. 
\end{acknowledgments}

\bibliography{ref}
\end{document}